\documentclass[3p,times,12pt]{elsarticle}

\usepackage{lineno,hyperref, amsmath, amssymb}

\journal{Nuclear Physics B}

\usepackage{graphicx}
\usepackage{color}
\usepackage{bm}
\usepackage{hyperref}
\usepackage{dcolumn}
\usepackage{slashed}

\newcommand{\ii}{\text{i}}
\newcommand{\dd}{\text{d}}
\newcommand{\be}{\begin{equation}}
\newcommand{\ee}{\end{equation}}
\newcommand{\bea}{\begin{eqnarray}}
\newcommand{\eea}{\end{eqnarray}}

\begin{document}

\begin{frontmatter}
\title{The quantum sine-Gordon model with quantum circuits}
\author[add_1]{Ananda Roy}
\ead{ananda.roy@tum.de}
\author[add_2]{Dirk Schuricht}
\author[add_3]{Johannes Hauschild}
\author[add_1,add_4]{Frank Pollmann}
\author[add_5]{Hubert Saleur}
\address[add_1]{Department of Physics, T42, Technische Universit\"at M\"unchen, 85748 Garching, Germany}
\address[add_2]{Institute for Theoretical Physics, Center for Extreme Matter and Emergent Phenomena, Utrecht University, Princetonplein 5, 3584 CE Utrecht, The Netherlands}
\address[add_3]{Department of Physics, University of California, Berkeley, CA 94720, USA}
\address[add_4]{Munich Center for Quantum Science and Technology (MCQST), 80799 Munich, Germany}
\address[add_5]{Institut de Physique Th\'eorique, Paris Saclay University, CEA, CNRS, F-91191 Gif-sur-Yvette}

\begin{abstract}
  Analog quantum simulation has the potential to be an indispensable technique in the investigation of complex quantum systems. In this work, we numerically investigate a one-dimensional, faithful, analog, quantum electronic circuit simulator built out of Josephson junctions for one of the paradigmatic models of an integrable quantum field theory: the quantum sine-Gordon (qSG) model in 1+1 space-time dimensions. We analyze the lattice model using the density matrix renormalization group technique and benchmark our numerical results with existing Bethe ansatz computations. Furthermore, we perform analytical form-factor calculations for the two-point correlation function of vertex operators, which closely agree with our numerical computations. Finally, we compute the entanglement spectrum of the qSG model. We compare our results with those obtained using the integrable lattice-regularization based on the quantum XYZ chain and show that the quantum circuit model is less susceptible to corrections to scaling compared to the XYZ chain. We provide numerical evidence that the parameters required to realize the qSG model are accessible with modern-day superconducting circuit technology, thus providing additional credence towards the viability of the latter platform for simulating strongly interacting quantum field theories.  
\end{abstract}
\end{frontmatter}

\section{Introduction}
The investigation of strongly interacting complex quantum systems remains one of the outstanding challenges of modern physics. Despite the remarkable progress on both numerical as well as analytical fronts, systematic and well-controlled non-perturbative analysis of many quantities of interest remain intractable. Quantum simulation provides a promising alternative to the aforementioned conventional techniques towards tackling these problems~\cite{Feynman_1982}. There are two approaches to quantum simulation: digital and analog. In principle, digital quantum simulation is universal~\cite{Lloyd1996, Lloyd1997}. It can be performed by a digital quantum computer built out of qubits. A digital quantum simulation of a many-body Hamiltonian comprises encoding of the target Hamiltonian as a trotterized sequence of one and two-qubit gates and readout of desired observables. However, opening up the digital quantum computer to externally applied gates and measurement apparatus leads to unavoidable decoherence of the physical qubits constituting the quantum computer. To combat for the finite lifetimes of the physical qubits as well as imperfections of the applied gate-set, quantum error-correction~\cite{Shor1995, Kitaev2003} is essential. In the recent years, spectacular process has been made towards realizing such a universal computing machine. In fact, noisy, intermediate-scale, quantum machines have already been shown to be capable of simulating certain aspects of few-body systems~\cite{Smith2019, Smith2019a}. However, extrapolating these efforts to the many-body domain will require enormous overhead and is likely to remain elusive in the immediate future. A more near-term approach to simulate many-body physics is analog quantum simulation, where a given  quantum system is tailored to simulate another~\cite{Doucot2004,Buchler2005,Cirac2010,Casanova2011, Houck2012, Roy2019}. In this approach the target Hamiltonian is realized by specifically engineering the given quantum system and letting it naturally evolve with time. There is no need to implement trotterized gate-sets. Thus, the physical degrees of freedom that comprise analog simulators do not need to be individually addressed and can be isolated from losses much more than their counterparts in a digital quantum computer. The imperfections in the engineered quantum system lead to `errors' which are typical of an experimental realization, {\it e.g.,} finite correlation lengths at criticality due to the disorder, etc. One of the biggest advantages of analog quantum simulation is the availability of a wide range of viable experimental platforms. Analog simulators based on trapped atoms have been used to experimentally simulate strongly correlated systems, topological phases of matter and gauge theories~\cite{Duca2015, Parsons2016, Bernien2017, Gross2017}, while trapped-ion based simulators have investigated problems in quantum magnetism~\cite{Friedenauer2008, Kim2010}. Finally, superconducting quantum electronic circuit (QEC) based simulators have been used to experimentally probe quantum electrodynamics in the strong and ultra-strong coupling regimes~\cite{Kuzmin2018, Kuzmin2019, Leger2019, PuertasMartinez2019}. 

In this work, we advance the research direction of faithful analog simulation of quantum field theories (QFTs) with QEC lattices~\cite{Roy2019}. Here, faithful refers to the fact that the degrees of freedom of the QFT are faithfully represented by the underlying lattice degrees of freedom and do not arise out of mathematical manipulations like bosonization. This is particularly important in multi-field QFTs, where properties as fundamental to a QFT as integrability, can be difficult to relate in the fermionic and the bosonic counterparts~\cite{Bukhvostov1980, Fateev1996, Lesage1997, Lesage1998, Saleur1999}. With this potential generalization to multi-field situations in mind, here we investigate, with the density matrix renormalization group (DMRG) technique, a faithful analog QEC simulator for one of the paradigmatic integrable QFT models: the quantum sine-Gordon~(qSG) model in 1+1 space-time dimensions. The QEC simulator is a one-dimensional array of suitably-arranged Josephson junctions and provides a faithful lattice-regularization for the qSG model using only local interactions. In the first part of the paper, we provide numerical evidence that the long-wavelength properties of the QEC lattice model are indeed described by the qSG field theory by computing various zero-temperature thermodynamic properties of the lattice model and comparing with analytical field-theory predictions. In the second part of the paper, we analyze the entanglement spectrum of the qSG model. To that end, we first compute the entanglement spectrum using the integrable Luther-Lukyanov lattice regularization~\cite{Luther1976,Lukyanov2003} involving  the quantum XYZ spin-chain~\cite{Baxter2013}. The entanglement spectrum of the XYZ chain is related to the spectrum of the corner-transfer-matrices (CTMs) of Baxter's eight-vertex model~\cite{Baxter2013} and thus, can be computed analytically~\cite{Ercolessi2009, Evangelisti2013}. Then, we calculate the entanglement spectrum of the QEC incarnation of the qSG model using DMRG. We show that the entanglement spectra of both the XYZ and the QEC regularizations comprise equidistant levels. We argue that the QEC and the XYZ entanglement level spacings are linearly related and verify this claim with numerical predictions. 

The analyzed QEC simulator provides a different lattice-regularization of the qSG model compared to the known existing ones. The primary motivation behind analyzing this lattice model is its experimental feasibility. It is a straightforward generalization of the current experimental works which have so far realized the free, compactified boson conformal field theory (CFT)~\cite{Kuzmin2018, Kuzmin2019, Leger2019, PuertasMartinez2019}. In fact, QEC systems which realize the qSG model in the semi-classcial limit have also  been fabricated and experimentally analyzed~\cite{Ustinov1995, Wallraff2003, Mazo2014}. However, as is shown in this work, the QEC regularization of the qSG model is also of intrinsic theoretical interest. In contrast to the XYZ chain regularization, where the qSG model arises out of Jordan-Wigner and bosonization transformations, in the QEC incarnation, the compact bosonic field $\phi$, is faithfully represented at the lattice level. This makes the QEC regularization more easily generalizable to multi-field scenarios~\cite{Roy2019}. At the same time, the correlation functions of vertex operators $e^{i\beta\phi}$, as computed using the QEC regularization, are less susceptible to corrections to scaling, where $\beta$ is the qSG coupling constant. These corrections arise when the correlation length of the underlying lattice model is not large enough compared to the lattice spacing; later we provide a quantitative estimate of when this effect becomes noticeable. The QEC model is more immune to the corrections to scaling compared to the XYZ chain since we start directly with bosonic fields on the lattice. This is in contrast to the XYZ chain, where spin-operator $\sigma^+$ is proportional to the qSG vertex operator $e^{i\beta\phi/2}$ to leading order~\cite{Lukyanov2003, Lukyanov1997, Lukyanov1999}. The situation for the XYZ model is further worsened by the fact that the qSG coupling and the mass of the soliton cannot be independently tuned unlike in the QEC lattice model. 
Additionally, we show that the same issues concerning reaching the scaling limit in the XYZ model also plague the entanglement spectrum -- for certain parameters, the XYZ model does not provide meaningful predictions for the qSG model, in contrast to the QEC model. Note that there is one other well-known regularization, proposed by Bogoliubov, Izergin and Korepin, of the qSG model based on nonlocal interactions of the bosonic degrees of freedom~\cite{Izergin1981, Korepin1993}. While the latter model is precious for analytical computations using the quantum inverse scattering method, its direct implementation in physical systems is not straightforward. Lastly, lattice regularizations based on QECs are amenable to DMRG and thus, provide a crucial tool to investigate properties of QFTs ({\it e.g.}, entanglement between spatially separated regions, etc) which are not easily accessible using alternative methods like the truncated conformal space approach~\cite{Yurov1990, Bajnok2002, Kuklijan2018}. 

The article is organized as follows. In Sec.~\ref{qSG_FT}, we briefly summarize the relevant properties of the qSG model. In Sec.~\ref{qec_model}, we describe the QEC lattice model and provide approximate expressions for the qSG parameters in terms of the lattice parameters. In Sec.~\ref{xyz_model}, we briefly recount the lattice-regularization based on the XYZ spin-chain. In Sec.~\ref{corr_fun}, we provide DMRG and analytical results for the one-point and two-point correlation functions. Secs.~\ref{qec_dmrg} and ~\ref{xyz_dmrg} describe the DMRG results for the QEC and XYZ regularizations respectively. The effects of corrections to scaling for the two models are discussed in Sec.~\ref{scale_correction}. Finally, in Sec.~\ref{ent_prop}, we compute the entanglement properties of the qSG model using DMRG. We first present the analytical and DMRG results for the XYZ model in Sec.~\ref{xyz_es}, followed by the DMRG results for the QEC lattice in Sec.~\ref{qec_es}. The consequences of the corrections to scaling on the entanglement spectrum for the two models are presented in Sec.~\ref{es_qec_vs_xyz}. Sec.~\ref{concl} presents a concluding summary and outlook. In~\ref{two_pt_fn}, we provide the analytical results of the zero-temperature two-point function for the qSG model.

\section{The quantum sine-Gordon model}
\label{qSG_FT}
In this section, we briefly summarize the basic properties of the qSG field theory that will be relevant for this work. More details on various properties can be obtained in, {\it e.g.}, Refs.~\cite{Coleman1975, Mandelstam1975, Zamolodchikov1979, Rajaraman1982, Bernard1991, Mussardo2010}. 

The qSG field theory is an integrable deformation of the free, compactified boson CFT. Its euclidean action is given by 
\begin{equation}
  S = \frac{1}{2}\int d^2x (\partial_\mu\phi)^2 + M_0\int d^2x\cos(\beta\phi),
\label{S_SG}
\end{equation}
where $M_0$ is the mass-parameter of the action and $\beta$ is the coupling constant. We set $\hbar=1$ throughout this work. We restrict ourselves to the regime when $\beta^2\in(0,8\pi)$. In the classical limit, which corresponds to $\beta\rightarrow0$, the coupling constant $\beta$ plays no role and can simply be scaled out. The resulting theory is the well-known classical sine-Gordon theory, which supports traveling wave-packet solutions, which propagate undistorted through the nonlinear wave-medium and scatter with only phase-shifts~\cite{Ablowitz2006}. Our interest, however, is in the quantum regime, when the parameter $\beta$ determines the spectrum of the theory. For $\beta<\sqrt{4\pi}$, the qSG model is in the attractive regime, where the spectrum of single-particle excitations consists of solitons, anti-solitons and breathers. For $\beta>\sqrt{4\pi}$, the model is in the repulsive regime when the spectrum of single-particle excitations consists only of solitons and antisolitons. The fermionized version of this model corresponds to the massive Thirring model~\cite{Coleman1975, Mandelstam1975, Thirring1958}, where the solitons are the fermions in the Thirring model. The choice $\beta = \sqrt{4\pi}$ corresponds to the free, massive (complex) fermion QFT. 

Many exact results are available for the qSG model, including its exact spectrum and the S-matrix. The following ones are relevant for our work.  The soliton mass can be derived to be~\cite{Zamolodchikov1995}
\begin{equation}
  M = \frac{2\Gamma\Big(\frac{\xi_{\rm SG}}{2}\Big)}{\sqrt{\pi}\Gamma\Big(\frac{1+\xi_{\rm SG}}{2}\Big)}\Bigg[\frac{M_0\pi \Gamma\Big(1-\frac{\beta^2}{8\pi}\Big)}{2\Gamma\Big(\frac{\beta^2}{8\pi}\Big)}\Bigg]^{\frac{1}{2-\frac{\beta^2}{4\pi}}},\ \xi_{\rm SG} = \frac{\beta^2}{8\pi - \beta^2}.  
  \label{Msol}
\end{equation}
The mass of the $n^{\rm{th}}$ breather state is given by 
\begin{equation}
  m_{n} = 2M\sin\frac{n\pi\xi_{\rm SG}}{2},\ n =1,2,\ldots,\Bigg\lfloor\frac{1}{\xi_{\rm SG}}\Bigg\rfloor.
  \label{Mb1}
\end{equation}
The ground state energy density with respect to the free, compactified boson CFT is~\cite{Zamolodchikov1995}
\begin{equation}
  E_0 (M) = -\frac{M^2}{4}\tan\frac{\pi\xi_{\rm SG}}{2}
\label{E_0_M}
\end{equation}
and the ground-state expectation value of the local vertex operator $e^{i\beta\phi}$ is given by~\cite{Lukyanov1996}  
\begin{equation}
 {\cal G}_\beta\equiv\langle e^{i\beta\phi}\rangle = \frac{(1+\xi_{\rm SG})\pi \Gamma\Big(1-\frac{\beta^2}{8\pi}\Big)}{16\sin(\pi\xi_{\rm SG})\Gamma\Big(\frac{\beta^2}{8\pi}\Big)}\Bigg[\frac{\Gamma\Big(\frac{1+\xi_{\rm SG}}{2}\Big)\Gamma\Big(1-\frac{\xi_{\rm SG}}{2}\Big)}{4\sqrt{\pi}}\Bigg]^{\frac{\beta^2}{4\pi} - 2}m_1^{\frac{\beta^2}{4\pi}}.
  \label{ebetaphi}
\end{equation}
Note that the last two predictions are for the continuum theory and in general, hold for the corresponding operators on the lattice only up to proportionality constants that depend on $\beta$; see, for example, similar computations for the XXZ model~\cite{Lukyanov2003} and the 2D classical Ising model in a magnetic field~\cite{Caselle1999}. Furthermore, we have used in the above expressions, the standard CFT normalization (note the difference in definition of $\beta$ compared to Ref.~\cite{Lukyanov1996}):
\begin{equation}
\langle \cos[\beta\phi(x)]\cos[\beta\phi(y)]\rangle\rightarrow \frac{1}{2}\frac{1}{|x-y|^{\frac{\beta^2}{2\pi}}},\ |x-y|\rightarrow0.
\end{equation}

\section{Lattice regularizations for the quantum sine-Gordon model}
In this section, we describe two lattice-regularizations for the qSG model. We start with the QEC lattice and discuss this in detail. Then, we briefly summarize the XYZ regularization following Ref.~\cite{Lukyanov2003}.

\subsection{QEC model}
\label{qec_model}
\begin{figure}
\centering
\includegraphics[width =\textwidth]{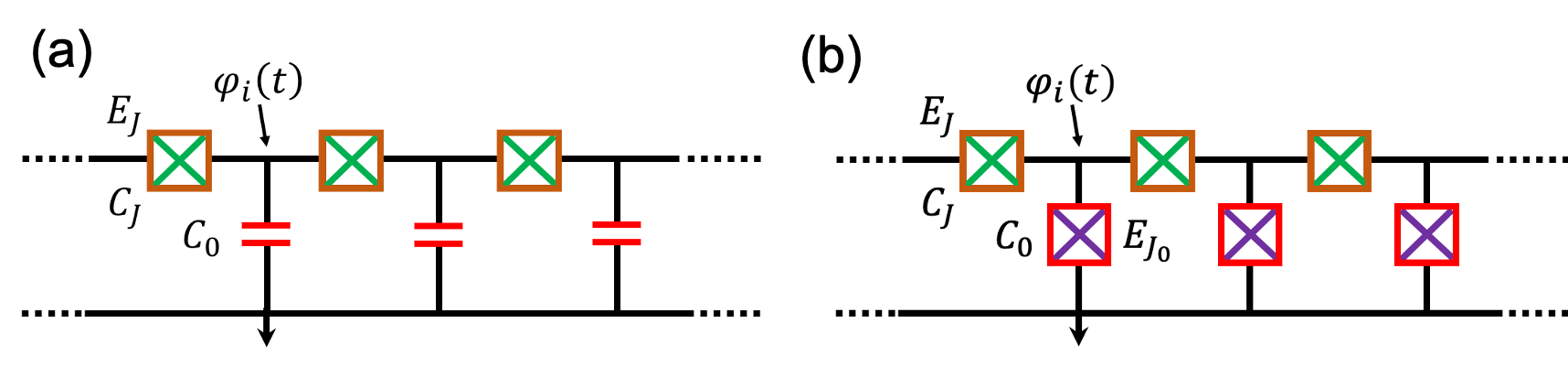}
\caption{\label{sine_gordon} Schematic of the 1D lattice-regularized model for the free, compactified boson CFT [panel (a)] and the qSG model [panel (b)]. Each Josephson junction is indicated by a cross with a box around it. (a) Each unit cell contains a Josephson junction (with junction energy $E_J$ and junction capacitance $C_J$) on the horizontal link, together with a capacitance to the ground-plane $C_0$.  We choose the parameter regime: $E_{C_J}\ll E_{C_0}\ll E_J$. In this limit, the nonlinearity of the Josephson junction on the horizontal link can neglected, giving rise to the free-compactified boson CFT. The kintetic inductance associated with the Josephson junction gives rise to Luttinger parameter $K\sim 1$. (b) In contrast to panel (a), there is a Josephson junction on the vertical link in each cell with junction energy $E_{J_0}$ and junction capacitance $C_0$. The desired parameter regime is $E_{C_J}\ll E_{C_0}, E_{J_0}\ll E_J$. The Josephson junction on the horizontal link leads to a $\beta^2\in(0,8\pi)$ [see Eq.~\eqref{beta_def}] while the Josephson junction on the vertical link gives rise to the cosine potential of the sine-Gordon model. In both cases, the bosonic field is the continuum version of the discretized superconducting phase $\varphi_i(t)$ at the node $i$ of the lattice. Physically, the $i^{\rm th}$ node denotes the $i^{\rm th}$ superconducting island. }
\end{figure}

The QEC lattice model for the qSG model comprises a 1D array of mesoscopic superconducting islands [see Fig.~\ref{sine_gordon}~(b)]. Two neighboring islands are separated by Josephson junctions (indicated by the green crosses within brown squares) with junction energy $E_J$ and charging energy $E_{C_J} = 2e^2/C_J$. In addition, each island is also separated from the ground plane by a Josephson junction (indicated by a purple cross within a red box), with junction energy $E_{J_0}$ and charging energy $E_{C_0} = 2e^2/C_0$. For reference, we show the QEC lattice for the free, compactified boson CFT in Fig.~\ref{sine_gordon}~(a). For the latter, the Josephson junction on the vertical link in each unit cell is replaced by a capacitance $C_0$. Throughout this work, we consider a homogeneous array with zero disorder in the off-set charges on the different superconducting islands.~\footnote{There are no qualitative changes for the qSG predictions in the experimentally relevant scenario of presence of imperfections, leading to disorder in the off-set charges; the free, compactified boson CFT has already been experimentally observed~\cite{Kuzmin2018, Kuzmin2019, Leger2019, PuertasMartinez2019}.} 
The Hamiltonian describing the array is given by
\begin{align}
\label{ham_arr}
H_{\rm{array}} = E_{C_0}\sum_{i=1}^Ln_i^2 + \delta E_{C_0}\sum_{i=1}^{L-1}n_in_{i+1} -E_g \sum_{i=1}^Ln_i- E_J\sum_{i=1}^{L-1}\cos(\varphi_i - \varphi_{i+1}) - E_{J_0}\sum_{i=1}^L\cos\varphi_i.
\end{align}
Here, the first term arises due to the finite charging energy of the mesoscopic islands and $n_i$ is the excess number of Cooper pairs on the $i^{\rm{th}}$ island.~\footnote{Note that $n_i$ can be both positive or negative, the latter corresponding to removal of a Cooper-pair from the superconducting condensate on the $i^{\rm th}$ island.} The finite junction capacitance $C_J$ leads to, in principle, infinite-range interaction between any two islands with a magnitude that decays exponentially with distance~\cite{Goldstein2013}. The relevant length scale is given by $a\sqrt{C_J/C_0}$, where $a$ is the lattice spacing. However, for realistic system parameters~\cite{Manucharyan2009}, it suffices to include only the nearest neighbor interaction~\cite{Glazman1997}, indicated by the second term in Eq.~\eqref{ham_arr} with $\delta$ being a small parameter $<1$. The third term arises due to the presence of a gate voltage $E_g$ at each node. 
The fourth term in Eq.~\eqref{ham_arr} describes the coherent tunneling of Cooper-pairs between neighboring islands. The last term describes the tunneling of Cooper pairs across the Josephson junctions on the vertical links and is responsible for the qSG nonlinearity. Note that the operators $n_i,\varphi_j$ are canonically conjugate satisfying $[n_i,e^{\pm i\varphi_j}] = \pm e^{\pm i\varphi_j}\delta_{ij}$. 

First, consider the case $E_{J_0}=0$ [this corresponds to Fig.~\ref{sine_gordon}~(a)]. Then, the QEC lattice realizes a version of the Bose-Hubbard model with nearest-neighbor repulsion~\cite{Glazman1997, Fisher1989, Fazio2001}, where the role of the bosons is played by Cooper-pairs. Since the number of excess Cooper-pairs can be both positive and negative, the model reduces to a Bose-Hubbard model for quantum rotors, with the number of Cooper-pairs being conserved. The phase-diagram of this model has been analyzed perturbatively~\cite{Glazman1997} and with DMRG~\cite{Roy2020a}. For $E_J\gg E_{C_0}$, the system is in a Luttinger liquid (LL) phase of Cooper-pairs, where its long wavelength properties are well-described by the action:\footnote{The estimate $E_J\gg E_{C_0}$ is perturbative. As shown below, quite small values of $E_J/E_{C_0}$ are sufficient to give rise to this phase.}
\begin{eqnarray}
S_0 = \frac{1}{2\pi K}\int d^2x \Big[\frac{1}{u}(\partial_t \varphi)^2 + u (\partial_x\varphi)^2\Big]
\end{eqnarray}
Here, $u$ is the plasmon velocity and $K$ the Luttinger parameter. Since the lattice model is non-integrable, exact expressions for $u,K$ are not known in terms of the lattice parameters. Perturbative estimates for $E_J\gg E_{C_0}$ are given by~\cite{Goldstein2013}
\begin{equation}
  \label{u_K}
u \simeq a\sqrt{2E_{C_0}E_J},\ K \simeq \frac{1}{2\pi}\sqrt{\frac{2E_{C_0}}{E_J}}.
\end{equation}
Lowering $E_J/E_{C_0}$ causes the system to transition into either a Mott-insulating or a charge-density-wave phase, with pinned densities $\rho = m/n$. In the Mott-insulating phases, $n = 1$ and $m\in\mathbb{Z}$, while for the charge-density-wave phases, $n =2$ and $m = 2k+1$, where $k\in\mathbb{Z}$. The transition out of the LL phase to either of the other two phases with fixed density $\rho_0 = m/n$ is caused by a perturbation of the form 
\begin{equation}
  S' \sim \int d^2x \cos[2n\theta + 2n\pi x(\rho - \rho_0)],
  \label{S_MI}
\end{equation}
where $\theta$ is the field dual to $\varphi$ and $\rho$ is the particle (Cooper-pair) density on each superconducting island. From dimensional analysis, it follows that the transition at fixed density $\rho=\rho_0$ occurs at a Luttinger parameter of $K_c = n^2/2$, while that with a change of density occurs at $K_c = n^2$~\cite{Giamarchi1997, Kuhner2000, Giamarchi2003}. For the model with only nearest-neighbor interactions, $n$ can be at most $2$. Thus, in the LL phase, the  Luttinger parameter $K\leq4$. 

It might appear surprising that we use Josephson junctions on the horizontal link, but consider only the limit $E_J\gg E_{C_J},E_{C_0}$, when the nonlinearity of the Josephson potential in the fourth term of Eq.~\eqref{ham_arr} plays no role. Indeed, we do not use the nonlinearity of the Josephson junctions on the horizontal links. However, the kinetic inductance associated with the Cooper-pairs, leads to a Luttinger parameter up to 4. If linear electromagnetic coil inductances were used instead of the Josephson junctions on the links of the array, apart from making the lattice boson non-compact, the Luttinger parameter would be $\sim Z/R_Q\sim0$, where $Z$ is the impendance of the array ($\sim 50\Omega$) and $R_Q$ is the impendance quantum ($\sim k\Omega$)~\cite{Manucharyan2009, Manucharyan2012}. As explained below, having $K\sim1$ is crucial for exploring the quantum regime of the qSG model. 

Now, consider the case $E_{J_0}\neq0$, which breaks the particle number conservation of the quantum rotor Bose-Hubbard model described above and contributes an additional term to the action:
\begin{eqnarray}
  S = S_0 + S_{\rm int},\ S_{\rm int} = M_0\int d^2x\cos\varphi(x). 
\end{eqnarray}
Rescaling the field $\varphi$ and the space-time axis: $t\rightarrow t\sqrt{u}$, $x\rightarrow x/\sqrt{u}$, we arrive at the action of the qSG model:
\begin{equation}
  S = \frac{1}{2}\int d^2x(\partial_\mu\phi)^2 + M_0\int d^2x\cos(\beta\phi),
  \label{S_SG_1}
\end{equation}
where $\varphi = \beta\phi$. Furthermore, the qSG coupling  and the mass-parameter of the action are
\begin{equation}
  \beta = \sqrt{\pi K}, \ M_0 = E_{J_0}a^{-\big(1-\frac{\beta^2}{4\pi}\big)}.
  \label{beta_def}
\end{equation}
Note that the free-fermion point of the qSG model occurs at $\beta=\sqrt{4\pi}$, which corresponds to $K=4$ and not $K=1$. From the above considerations, thus, the QEC lattice model of Eq.~\eqref{ham_arr} is limited to the attractive regime of the qSG model: $\beta\leq \sqrt{4\pi}$ since it has only nearest-neighbor interactions. However, in an actual experimental realization, the interaction can, in principle, be long-ranged, but exponentially decaying. The range of the interaction is determined by the ratio $C_0/C_J$~\cite{Goldstein2013}. The longer-range model, which supports larger Luttinger parameters, is important for realizing the repulsive regime of the qSG model and can be similarly analyzed generalizing this current work. 

Note two important features of the QEC lattice model for the qSG model. First, the underlying lattice degrees of freedom accessible to numerical simulations are the vertex operators $e^{i\varphi_i}$, which, to leading order, up to a proportionality constant depending on $\beta$, are the same as the QFT vertex operators $e^{i\beta\phi}$. Second, the deviations from the leading qSG action are captured by the term given in Eq.~\eqref{S_MI}. They are irrelevant in the renormalization group sense, but more importantly, do not renormalize the relevant mass term in the qSG action. This allows us to have an analytical control on the mass-parameter of the continuum qSG action in terms of the QEC lattice parameters.

\subsection{XYZ model}
\label{xyz_model}
Now, we briefly summarize the qSG limit of the XYZ chain following Ref.~\cite{Lukyanov2003}. Consider the Hamiltonian:
\begin{equation}
  \label{ham_xyz}
H_{\rm XYZ} = -\frac{1}{2}\sum_{i=1}^{L-1}\Big[J_x\sigma_i^x\sigma_{i+1}^x +J_y\sigma_i^y\sigma_{i+1}^y + J_z\sigma_i^z\sigma_{i+1}^z\Big],
\end{equation}
where $J_{x,y,z}$ are the coupling constants and $\sigma_{x,y,z}$ are Pauli-matrices. We consider the parameter regime: $J_x\geq J_y\geq|J_z|$.  For $J_x=J_y$, the excitation spectrum is gapless and the low-energy properties of the model are described by a Luttinger liquid action~\cite{Giamarchi2003}. The correlation length $\xi_{\rm{XYZ}}$ diverges as
\begin{equation}
\label{xi_xyz}
\xi_{\rm XYZ} \simeq \frac{1}{4}\Bigg(\frac{4}{l}\Bigg)^{\frac{1}{1-\frac{K_{\rm XYZ}}{2}}}, \ l^2 = \frac{J_x^2-J_y^2}{J_x^2 - J_z^2}, \ K_{\rm XYZ} = \frac{2}{\pi}\cos^{-1}\frac{J_z}{J_x},
\end{equation}
where we denote the Luttinger parameter of the spin-chain by $K_{\rm XYZ}$. Close to the critical point, the system is described by the qSG field theory, with the action of Eq.~\eqref{S_SG_1}. In this case, the qSG coupling and the mass of the soliton are given by
\begin{equation}
  \label{k_xyz}
\beta^2 = 4\pi K_{\rm XYZ},\ M = \frac{1}{a\xi_{\rm XYZ}},
\end{equation}
where $a$ is the lattice-spacing. For the continuum theory to be applicable, it is necessary to have $\xi_{\rm XYZ}\gg 1$, which in turn restricts the achievable values of M. Note that for the XYZ chain, the free-fermion point of the qSG model occurs at $K_{\rm XYZ} = 1$. To leading order, the spin-creation operator $\sigma^\pm(x)$ can be identified with the vertex-operator of the qSG model~\cite{Lukyanov2003}:
\begin{equation}
\label{sp_vert}
\sigma^\pm(x)\simeq C_0e^{\pm \frac{i\beta\phi(x)}{2}} + \ldots, \ C_0 = \frac{1}{2\big(1-\frac{\beta^2}{8\pi}\big)\sqrt{Z_{1,0}}}\Bigg(\frac{M}{4}\Bigg)^{\frac{\beta^2}{16\pi}},\ \sqrt{Z_{1,0}} = a^{\frac{\beta^2}{16\pi}}\langle e^{\frac{i\beta\phi}{2}}\rangle, 
\end{equation}
where the dots indicate corrections to scaling arising from irrelevant terms. More details on the corrections to scaling and their effect on the correlation functions in the scaling limit of the XYZ chain can be found in Ref.~\cite{Lukyanov2003}.

Note that, as for the QEC model, the effective action is not just the qSG action of Eq.~\eqref{S_SG_1}, but include corrections, which are irrelevant in the renormalization group sense (see Ref.~\cite{Lukyanov2003} for explicit forms for these correction terms). However, in contrast to the QEC model, the local spin operator corresponding to the vertex-operator accessible in the XYZ case is $e^{\frac{i\beta\phi}{2}}$, which is semi-local. As will be shown later, the expectation value of this vertex operator as well as entanglement properties of the qSG limit of the XYZ chain are more susceptible to corrections to scaling compared to that obtained from the QEC lattice model. 

\section{Zero-temperature computations of correlation functions}
\label{corr_fun}
In this section, we compute various zero-temperature thermodynamic properties of the two different lattice regularizations of the qSG model using DMRG and compare with analytical predictions for the qSG field theory. The DMRG computations were performed using the TeNPy package~\cite{Hauschild2018}.
\subsection{QEC model}
\label{qec_dmrg}
First, we present the DMRG results for the QEC lattice model. The local Hilbert space on each island was truncated to 9 dimensions: $n_i  = -4,-3,\ldots,3,4$ (similar computations for the free, compactified boson CFT, including the evidence that this truncation of the local Hilbert space is sufficient, are done in Refs.~\cite{Roy2020a, Roy2020}). For definiteness, we chose $\delta=0.2$ and set the overall energy scale by choosing $E_{C_0} = 1$ for all the computations of the QEC model.

\begin{figure}
  \centering
  \includegraphics[width = 0.95\textwidth]{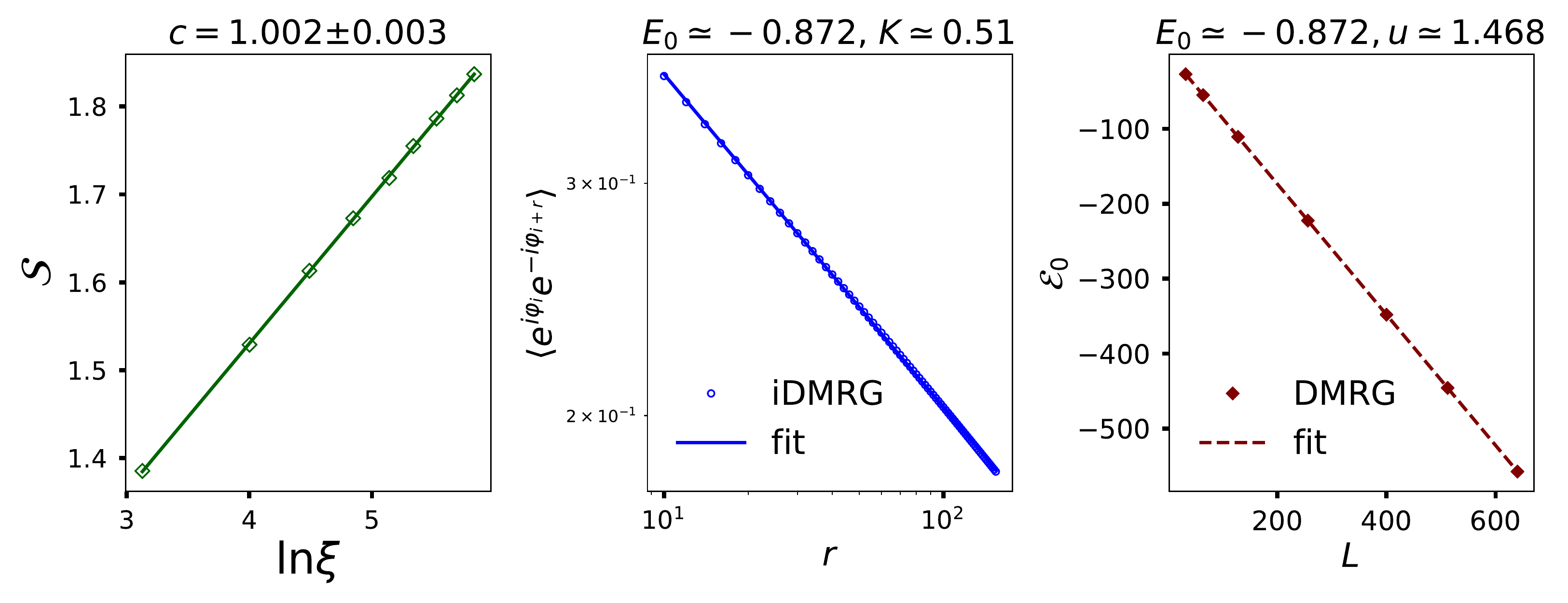}
  \caption{\label{fig:u_K_1} Characteristics of the LL phase obtained using DMRG for the QEC model. We chose $E_J/E_{C_0} = 1.55$. Recall that $E_{J_0} = 0$. (Left) From the logarithmic growth of the entanglement entropy ${\cal S}$ as a function of correlation length $\xi$ [see Eq.~\eqref{ent_entropy_form}], we obtain the central charge $c$. (Center) The algebraic decay of correlations $\langle e^{i\varphi_i}e^{-i\varphi_{i+r}}\rangle $ as a function of $r$ in log-log scale [see Eq.~\eqref{K_form}] obtained using infinite DMRG. The extracted Luttinger parameter as well as the computed ground state energy density are shown. (Right) Variation of the ground-state energy as a function of the system size obtained using finite DMRG. A fit to Eq.~\eqref{u_form} using the obtained value of $c$ yields $E_0$ and $u$. As shown, the value of $E_0$ matches that obtained using infinite DMRG to the third decimal place.  }
\end{figure}

\begin{figure}
  \centering
  \includegraphics[width = 0.85\textwidth]{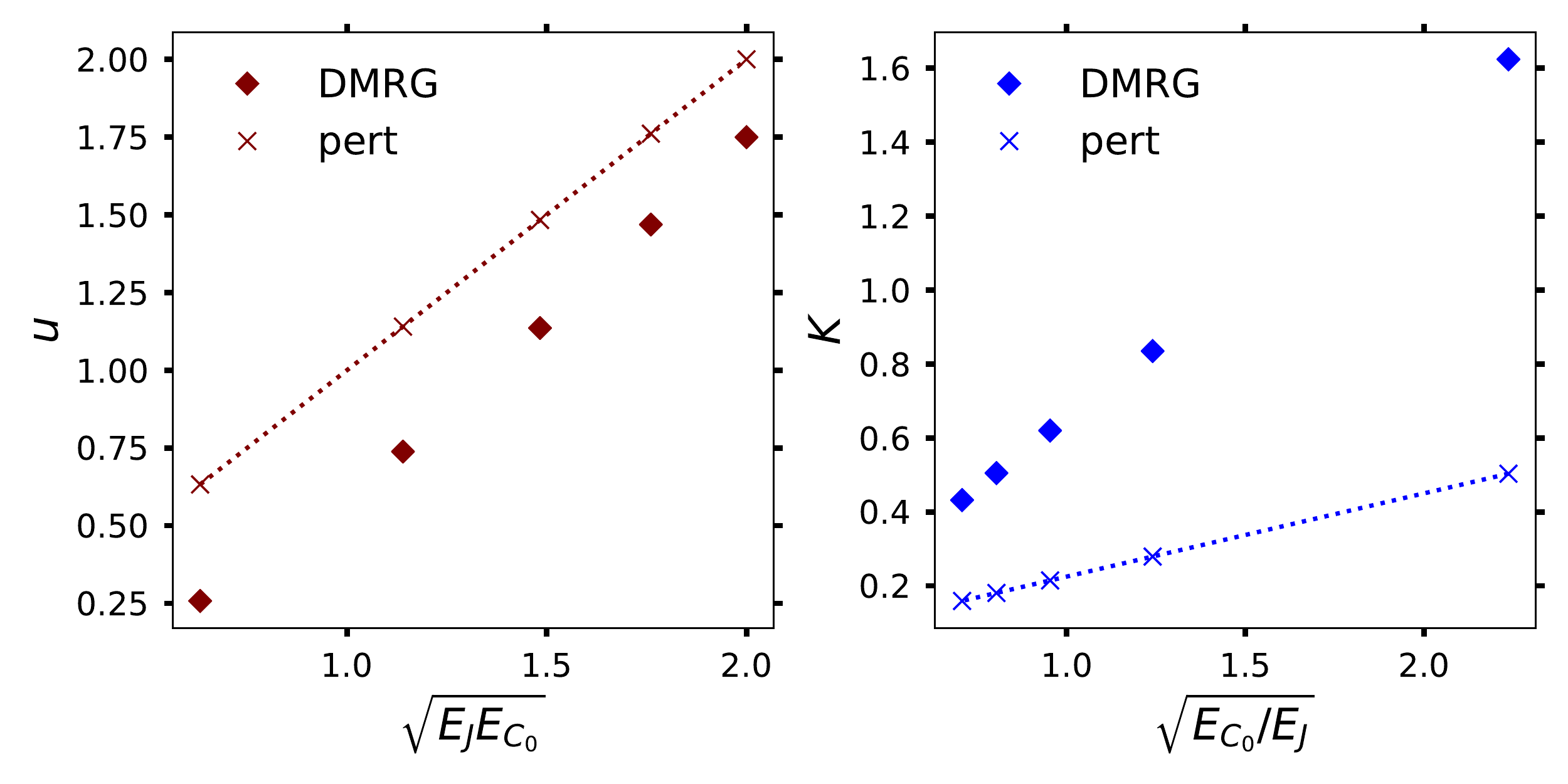}
  \caption{\label{fig:u_K_2} DMRG results (solid diamonds) for the variation of the plasmon velocity, $u$ (left panel),  and the Luttinger parameter, $K$ (right panel), with $E_J/E_{C_0}$ in the LL phase ($E_{J_0} = 0$). The corresponding values predicted by the approximate analytical formula [Eq.~\eqref{u_K}] are shown with crosses. Although as $E_J$ increases, the sets of values approach each other, for the parameters considered in this work, as shown in this figure, the perturbative formulas are not accurate. }
\end{figure}

First, we consider the case $E_{J_0} = 0$, when the system is described by a free compactified, boson CFT. The three key properties of this CFT are its central charge ($c$), the compactification radius $R = 1/\sqrt{\pi K}$ and the plasmon velocity $u$. They can be obtained from DMRG computations in the following way. The central charge can be verified by computing the scaling of the entanglement entropy (${\cal S}$) with correlation length ($\xi$) for a partitioning of an infinite system into two semi-infinite halves\footnote{This behavior is also true for gapped systems and is not a characteristic of the gapless spectrum. Here, we perform a `finite entanglement scaling'~\cite{Pollmann2009}, where successive increase of the DMRG bond-dimension increases the correlation length. }~\cite{Pollmann2009, Calabrese2004, Tagliacozzo2008, Calabrese2009}: 
\begin{equation}
  {\cal S} = \frac{c}{6}\ln\xi.
  \label{ent_entropy_form}
\end{equation}
The Luttinger parameter of the theory is obtained by computing the algebraic decay of the correlation function~\cite{Kuhner2000, Giamarchi2003}:
\begin{equation}
\langle e^{i\varphi_i}e^{-i\varphi_{i+r}}\rangle \sim \frac{1}{|r|^{K/2}}. 
  \label{K_form}
\end{equation}
Finally, the plasmon velocity $u$ is obtained by computing the zero-temperature ground-state energy (${\cal E}_0$) for a finite-system of size $L$ with open (free) boundary conditions and fitting to the Cardy formula~\cite{Cardy1989, diFrancesco1997}:
\begin{equation}
  {\cal E}_0 = E_0 L  - \frac{\pi c u}{24L}, 
  \label{u_form}
\end{equation}
where $E_0L$ is the extensive contribution to the ground state energy. The results are shown in Fig.~\ref{fig:u_K_1} for $E_J/E_{C_0} = 1.55$. The central charge and the Luttinger parameter are extracted using infinite DMRG (left and center panels). We also obtain the ground state energy density $E_0$ from the infinite system simulations. The right panel of Fig.~\ref{fig:u_K_1} shows the variation of the ground state energy as a function of system size for free boundary conditions obtained using finite DMRG. By fitting to Eq.~\eqref{u_form} and using the obtained value of the central charge, we get the ground-state energy density $E_0$ and the plasmon velocity $u$. As shown, both the finite and infinite DMRG results for $E_0$ match to the third decimal place. 
The extracted plasmon velocity and the Luttinger parameter values as $E_J/E_{C_0}$ is varied are shown with solid diamonds in Fig.~\ref{fig:u_K_2}. The corresponding perturbative analytical estimates, from Eq.~\eqref{u_K}, are shown with crosses. While the analytical and the DMRG results approach each other for large $E_J/E_{C_0}$, for the parameters of interest in this work, the perturbative analytical estimates are clearly insufficient. 

\begin{figure}
  \centering
  \includegraphics[width = 0.65\textwidth]{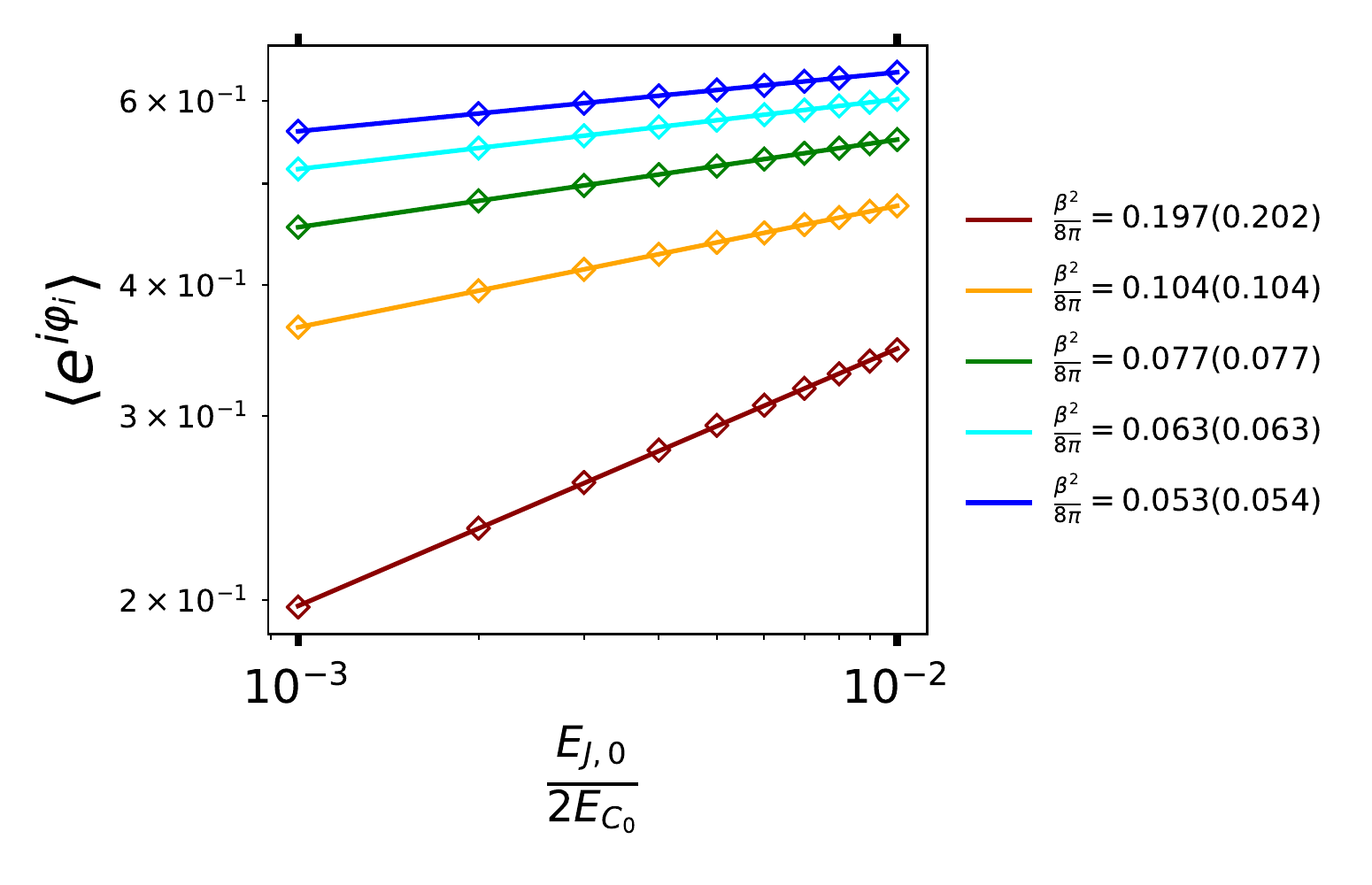}
  \caption{\label{fig:ephi} Variation of $\langle e^{i\varphi_i}\rangle$, obtained using infinite DMRG, as the mass-parameter of the qSG action, determined by $E_{J_0}/E_{C_0}$, is changed for $E_J/E_{C_0}$ $=0.2, 0.55, 1.1, 1.55$ and $2$. Here, $i$ is a site within an infinitely large QEC lattice. For these choices of $E_J/E_{C_0}$, the obtained Luttinger parameters are shown in Fig.~\ref{fig:u_K_2}. The Luttinger parameter determines the qSG coupling $\beta$ through Eq.~\eqref{beta_def}. The value of $\beta^2/8\pi$ computed using $K$ is shown in parenthesis in the legend for each curve. The algebraic dependence predicted in  Eq.~\eqref{ebetaphi} is verified, with the slope giving $\beta^2/(8\pi - \beta^2)$. The value of $\beta^2/8\pi$ obtained from the slopes is shown in the legend. As is evident, the agreement is quite good and improves as $E_J/E_{C_0}$ increases. This is because increasing $E_J/E_{C_0}$ makes the irrelevant corrections to the scaling field theory action, given in Eq.~\eqref{S_MI}, less dominant. Then, the qSG description is better suited. }
\end{figure}

Now, we consider the case when $E_{J_0}\neq0$. First, we compute the expectation value of the local lattice vertex operator $\langle e^{i\varphi_i}\rangle$ and compare with the QFT predictions for the continuum vertex operator $\langle e^{i\beta\phi}\rangle$, given in Eq.~\eqref{ebetaphi}. Here $i$ is a lattice point within an infinitely large chain. The results obtained with infinite DMRG are shown in Fig.~\ref{fig:ephi}. For a given choice  $E_J/E_{C_0}$, which fixes the Luttinger parameter and thus, $\beta$ [see Eq.~\eqref{beta_def}], we compute $\langle e^{i\varphi_i}\rangle$ as a function of $E_{J_0}/2E_{C_0} = M_0/2$. We verify the expected algebraic dependence [see Eqs.~(\ref{Msol}, \ref{Mb1}, \ref{ebetaphi})] for different choices of $E_J/E_{C_0}$ (note the log-log scale for the plot).  For each choice of $E_J/E_{C_0}$, the slope is $\beta^2/(8\pi - \beta^2)$. The latter can be used to compute the value of $\beta^2/8\pi$ which is shown in the legend of the plot. In parenthesis, for each curve, the expected value of $\beta^2/8\pi$ is shown obtained by computing the Luttinger parameter for $E_{J_0}/E_{C_0}=0$ [see Fig.~\ref{fig:u_K_2}]. The agreement is reasonably good, and improves as $E_J/E_{C_0}$ is increased. This improvement is because the corrections to the qSG description of the lattice model, shown in Eq.~\eqref{S_MI}, become more irrelevant in the renormalization group sense. 

\begin{figure}
  \centering
  \includegraphics[width = 0.5\textwidth]{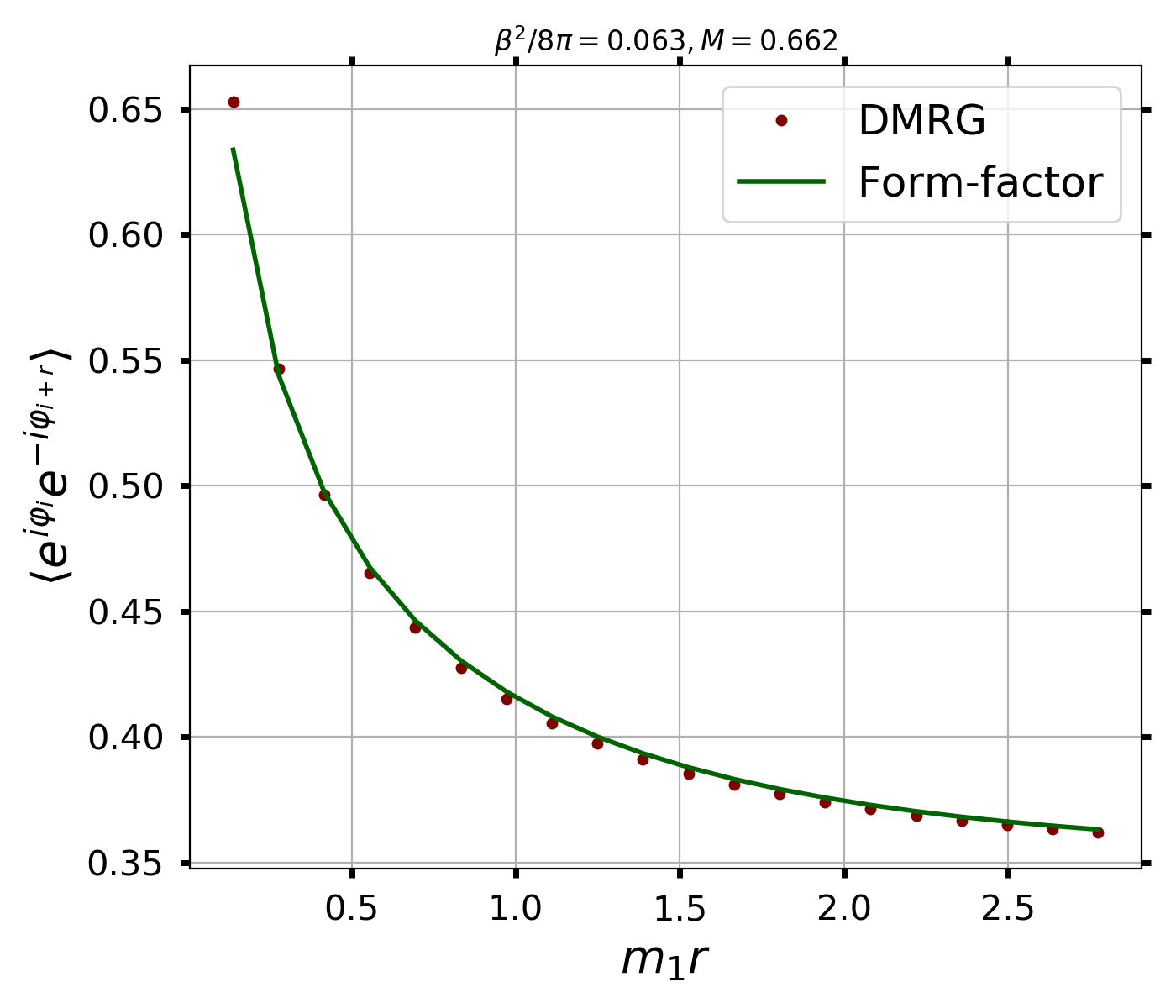}
  \caption{\label{fig:correln} Two-point correlation function $\langle e^{i\varphi_i}e^{-i\varphi_{i+r}}\rangle$ as function of $r$ computed using infinite DMRG (purple dots). We chose  $E_J/E_{C_0}=1.55$ ({\it i.e.,} $\beta^2/8\pi = 0.063$) and $M_0/2=E_{J_0}/E_{C_0} = 0.016$. The corresponding soliton mass ($M$), determined using Eq.~\eqref{Msol} is $\simeq0.662$. The plasmon velocity ($u\simeq1.46$) is determined using the Casimir energy as in Fig.~\ref{fig:u_K_2}. The corresponding form-factor results are shown with  the solid green line. The form-factor computations are performed including up to the second breather mass [see \ref{two_pt_fn}]. The overall field normalization to relate the lattice operators to the continuum ones is determined by computing the one-point functional (Fig.~\ref{fig:ephi}). Note that there are no fit parameters in this plot. }
\end{figure}

Note that the above computation only confirms the overall scaling of the expectation value of the local field, which could be inferred purely from dimensional analysis. However, the exact magnitudes of the lattice and continuum vertex operators are equal only up to a proportionality factor, which depends on $\beta$ or equivalently $E_J/E_{C_0}$. This proportionality constant arises since the predictions for the qSG field theory assume a certain normalization for the correlation functions of the vertex operators in the conformal limit.~\footnote{The proportionality constant is expected to have a certain scaling behavior, which could be analyzed further; similar results exist for the 2D Ising model in a magnetic field~\cite{Caselle1999}.} We determine this constant of proportionality and use this to compute the two-point correlations of the vertex operators: $\langle e^{i\beta\phi(0)}e^{-i\beta\phi(r)}\rangle$. The corresponding operators to be computed using DMRG are $\langle e^{i\varphi_i}e^{-i\varphi_{i+r}}\rangle$. Using the corresponding form-factors for vertex operators~\cite{Smirnov1992, Lukyanov1997a}, we compute the relevant two-point correlation function. The form-factor computation is performed by including contributions up to second breather mass (see~\ref{two_pt_fn} for more details). Here we only present the comparison between the DMRG and the analytical results in Fig.~\ref{fig:correln}. We chose  $E_J/E_{C_0}=1.55$ ({\it i.e.,} $\beta^2/8\pi = 0.063$) and $E_{J_0}/E_{C_0} = 0.016$ [see also Eq.~\eqref{beta_def}]. The corresponding soliton mass, determined using Eq.~\eqref{Msol} is $\simeq0.662$. The plasmon velocity determined using the Casimir energy is $\simeq 1.46$, see Fig.~\ref{fig:u_K_2}. The overall field normalization is determined using by computing the expectation value of the one-point expectation value (see Fig.~\ref{fig:ephi}). Note that there are no fit parameters in this plot since the mass of the soliton is determined analytically and the field normalization is determined from a different, independent computation. Similar results were obtained for other choices of $E_J/E_{C_0}$ and are not shown for brevity. The infinite DMRG computations are shown with maroon dots, while the form-factor predictions with a solid green line.  

\subsection{XYZ model}
\label{xyz_dmrg}
\begin{figure}
  \centering
  \includegraphics[width = \textwidth]{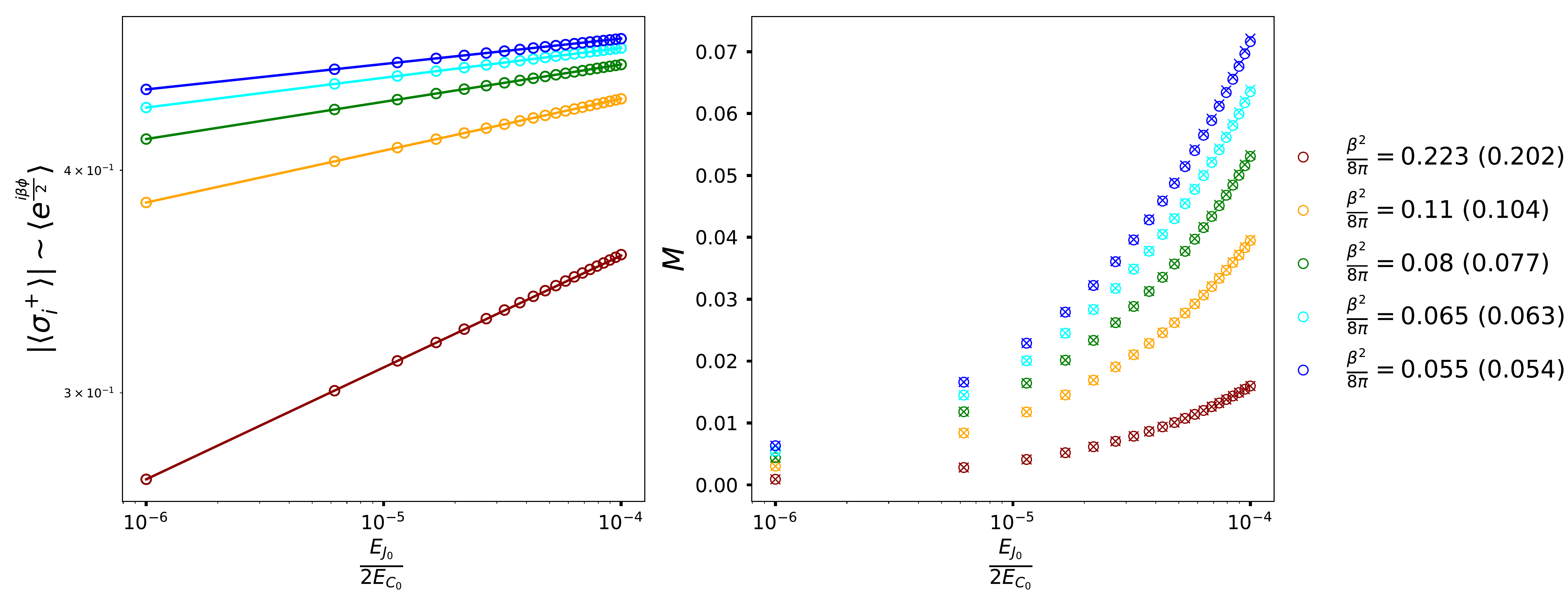}
  \caption{\label{fig:ebetaphi2} Variation of the spin operator ($\langle\sigma_i^+\rangle$ $\sim$ $\langle e^{i\beta\phi/2}\rangle$, left panel) and the soliton mass ($M$, right panel) as a function of the mass parameter of the qSG action, $M_0=E_{J_0}/2E_{C_0}$ [see Eqs.~(\ref{S_SG_1},\ref{beta_def})]. The different colors correspond to the different choices of $\beta^2/8\pi$ (chosen to be the same set as in Fig.~\ref{fig:ephi}). From the slope, the extracted values of the $\beta^2/8\pi$ [see Eq.~\eqref{ebetaphi2}] are shown in the legend, together with the expected values in parenthesis. The soliton mass is obtained from the  infinite DMRG results for $\langle \sigma_i^+\rangle$ using Eq.~\eqref{sp_vert}. These are shown on the right panel with empty circles for the different choices of $\beta^2/8\pi$. The corresponding analytical predictions [Eq.~\eqref{k_xyz}] are shown with crosses. Notice that the x-axis values in both panels is much smaller compared to Fig.~\ref{fig:ephi}. This is because the XYZ chain is no longer in the scaling regime for the range of $E_{J_0}/E_{C_0}$ shown in Fig.~\ref{fig:ephi}. This is discussed in Sec.~\ref{scale_correction}.}
\end{figure}

In this section, we present DMRG results for the qSG limit of the XYZ chain. We choose the same set of $\beta$ as in Sec.~\ref{qec_dmrg} and vary the mass-parameter of the qSG action $M_0$ [see Eq.~\eqref{S_SG_1}]. To make comparison with the QEC lattice model easier, we plot the results with respect to the corresponding QEC mass parameter $E_{J_0}/E_{C_0}$ [see  Eq.~\eqref{beta_def}]. For the simulations, we set $J_x=1$. From Eqs.~(\ref{Msol}, \ref{xi_xyz}, \ref{k_xyz}), the corresponding values of $J_y,J_z$ are inferred. The results obtained with infinite DMRG are shown in Fig.~\ref{fig:ebetaphi2}. In the left panel, we plot $\langle \sigma_i^+\rangle$ $\sim$ $\langle e^{i\beta\phi/2}\rangle$ as a function of $M_0/2 = E_{J_0}/2E_{C_0}$ on a log-log scale. This expectation value scales as~\cite{Lukyanov1996}
\begin{equation}
  \langle \sigma_i^+\rangle\sim\langle e^{i\beta\phi/2}\rangle\sim \Bigg(\frac{E_{J_0}}{E_{C_0}}\Bigg)^{\frac{\beta^2/4}{8\pi - \beta^2}}.
  \label{ebetaphi2}
\end{equation}
The proportionality constant of the second relationship is also conjectured~\cite{Lukyanov1996}, but the exponent of the algebraic dependence is sufficient for our purposes. The extracted values of $\beta^2/8\pi$ are shown for the different data points with the expected values within parenthesis. As an extra check, we compute also the soliton mass ($M$) as a function of $E_{J_0}/E_{C_0}$. This is directly computed from Eq.~\eqref{sp_vert} by eliminating $\langle e^{i\beta\phi/2}\rangle$. The results obtained with infinite DMRG are shown on the right panel (empty circles), together with the analytical values of Eq.~\eqref{k_xyz} (crosses). As is evident, the agreement is reasonable, which gives us confidence that we are indeed probing the qSG regime of the XYZ parameter space. 

Note that the x-axis values for both the panels in Fig.~\ref{fig:ebetaphi2} are several orders of magnitude smaller than that of the QEC simulations of Fig.~\ref{fig:ephi}. This is because for the values of the QEC simulations, the XYZ chain is no longer in the scaling limit. The reader will notice that the last few data points in the plot of the soliton mass for $\beta^2/8\pi = 0.054$ (in blue) already start showing deviations occurring due to corrections to scaling. We discuss this next. 

\subsection{Corrections to scaling: QEC vs XYZ}
\label{scale_correction}
So far, we have computed various zero-temperature thermodynamic quantities for the qSG field theory using DMRG for the QEC and the XYZ lattice regularizations. Now, we discuss how the two regularizations fare with regards to corrections to scaling. Consider a lattice model whose long wavelength behavior ({\it i.e.}, the scaling limit) is governed by a QFT. Then, there are two corrections to the scaling-limit behavior. First, the lattice Hamiltonian gives rise to various terms in addition to the QFT Hamiltonian, which are irrelevant in the renormalization group sense. The contributions of these terms are small, but nonzero. Second, the QFT operators are approximately represented by the lattice operators. Thus, while computing correlation functions of QFT operators using lattice regularizations, both these corrections contribute to the subleading corrections.
For the QEC incarnation of the qSG model, the corrections to the scaling-limit action include the irrelevant term shown in Eq.~\eqref{S_MI}, as well as irrelevant terms formed by higher descendants of the vertex operator $e^{i\beta\phi}$. A similar set of corrections arise for the effective Hamiltonian of the XYZ chain regularization -- some of these terms are explicitly computed in Ref.~\cite{Lukyanov2003}. However, the two regularizations behave differently when it comes to the definitions of the vertex operators whose correlations are computed. This is because while the spin-operator $\sigma_i^+$ of the XYZ chain is approximately equal to $e^{i\beta\phi/2}$ [see Eq.~\eqref{sp_vert}], the lattice operator $e^{i\varphi_i}$ of the QEC lattice model, to leading order, is equal to the operator $e^{i\beta\phi}$ up to an overall $\beta$-dependent proportionality constant.~\footnote{In this discussion, we have concentrated on the qSG vertex operators whose correlation functions are observable by measuring, for instance, current-current correlation functions in a QEC experiment. However, this does not preclude the existence of local operators which have equal or better resilience to corrections to scaling in the XYZ model. Such operators are likely to be superpositions of local qSG operators specifically chosen to be local in the spin language.} 

We observe numerically that larger correlation lengths, $\xi_{\rm XYZ}$, (equivalently, smaller soliton mass, $M$), are required to reach the scaling regime of the XYZ spin-chain compared to the QEC lattice. However, this is not always possible in practice due to the following. From Eq.~\eqref{xi_xyz}, for $\xi_{\rm XYZ}\rightarrow\infty$, the parameter $l\rightarrow0$. This is accomplished by choosing $|J_x^2 - J_y^2|\ll |J_x^2 - J_z^2|$. However, as $\beta^2$ gets close to either 0 or $8\pi$, this becomes increasingly difficult since $|J_z/J_x|\rightarrow1$ for these choices. This is the other crucial difference between the QEC and the XYZ regularizations. {\it The QEC regularization allows the two qSG parameters: $\beta, M_0$ to be independently controlled}: the first being controlled by the junction energies ($E_J$) of the Josephson junctions on the horizontal links, while the latter being controlled by the that ($E_{J_0}$) of the Josephson junctions on the vertical links. In contrast, {\it in the XYZ regularization, $M_0$ cannot be tuned independent of $\beta$}. 

\begin{figure}
  \centering
  \includegraphics[width = \textwidth]{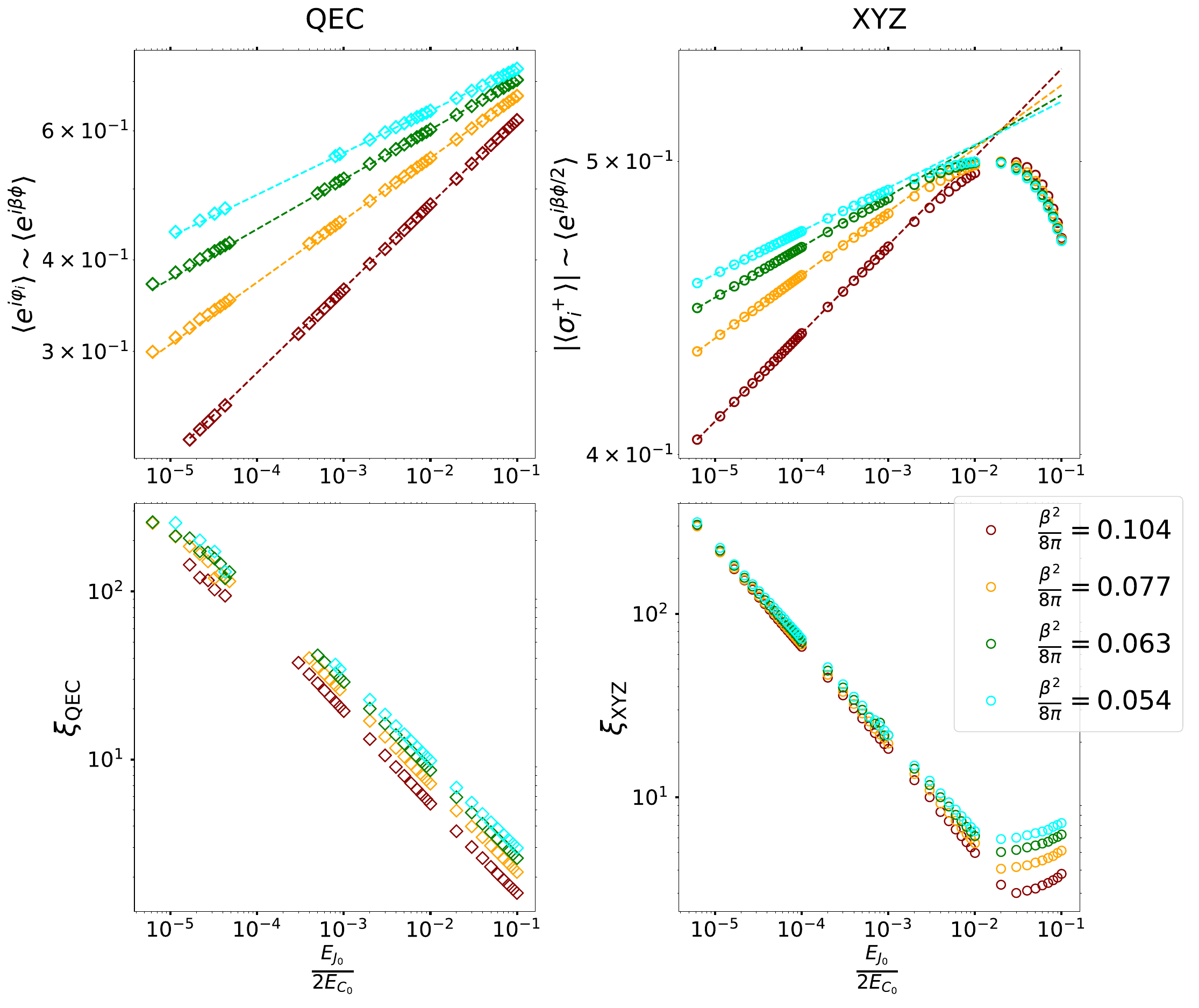}
  \caption{\label{fig:ebetaphi_XYZ_QEC} Comparison of the expectation values of the vertex operators obtained using infinite DMRG for the QEC lattice model ($\langle e^{i\varphi_i}\rangle\sim\langle e^{i\beta\phi}\rangle$, top left) and the XYZ chain ($\langle\sigma_i^+\rangle\sim \langle e^{i\beta\phi/2}\rangle$, top right) as a function of $M_0/2 \propto E_{J_0}/2E_{C_0}$, the mass-parameter of the qSG field theory [see Eqs.~(\ref{S_SG_1}, \ref{beta_def})]. The field theory computations for both the quantities predict a linear dependence (see Figs.~\ref{fig:ephi}, \ref{fig:ebetaphi2}). For very small $M_0/2$, where the correlation-length is large for both QEC and the XYZ lattice models and the continuum qSG description is valid for both models. As $E_{J_0}/E_{C_0}$ increases, the correlation length diminishes and the corrections to scaling becomes important. However, as seen from the QEC plot, the expectation value of the vertex operator continues to follow the straight line, while that from the XYZ chain deviates from the field theory predictions at $E_{J_0}/E_{C_0}\sim10^{-3}$. This difficulty of reaching the scaling regime for the XYZ chain occurs in the region where $|J_x - J_y|$ is no longer $\ll |J_x - J_z|$. Then $l$ no longer tends to zero and thus, the correlation length $\xi_{\rm XYZ}$ is no longer large [see Eq.~\eqref{xi_xyz}]. We do not go beyond $E_{J_0}/2E_{C_0}=0.1$ since the correlation length for both the XYZ and the QEC models is only a few lattice sites and it is not meaningful to apply the qSG field theory predictions beyond this point. For reference, the corresponding correlation lengths are shown in the bottom left and bottom right panels. Note that the correlation length for the XYZ chain actually goes up for $E_{J_0}/E_{C_0}>10^{-2}$. This is because for this choice of parameters, the XYZ chain is no longer in the scaling limit and increasing the aforementioned ratio no longer corresponds to increasing the mass gap of the model (see main text for more details).}
\end{figure}

This difference in corrections to the scaling limits for the two regularizations is shown in Fig.~\ref{fig:ebetaphi_XYZ_QEC}. The top left and top right panels show the variation of the expectation values of local fields as a function of the qSG mass-parameter $M_0/2 \propto E_{J_0}/2E_{C_0}$ [see Eqs.~(\ref{S_SG_1}, \ref{beta_def})]. The QEC lattice results for $\langle e^{i\varphi_i}\rangle\sim\langle e^{i\beta\phi}\rangle$ and the XYZ results for $\langle \sigma_i^+\rangle\sim\langle e^{i\beta\phi/2}\rangle$ are shown. The qSG field theory predicts a linear-dependence with $M_0$ for both vertex operators on a log-log scale [see Eqs.(\ref{ebetaphi}, \ref{sp_vert}, \ref{ebetaphi2})]. As seen from Fig.~\ref{fig:ebetaphi_XYZ_QEC}, the QEC vertex operator exhibits this behavior for {\it all} choices of $M_0$. On the other hand, the corresponding XYZ spin operator does so only deep in the scaling regime when $M_0< 10^{-3}$ for the shown choices of $\beta^2/8\pi$. One can identify the region where the corrections to scaling are noticeable as the region where $|J_x - J_y|$ is no longer $\ll |J_x - J_z|$. In this region, $l$ is no longer small and thus, the correlation length $\xi_{\rm XYZ}$ is no longer large [Eq.~\eqref{xi_xyz}]. Note that the problem is more severe for $\beta^2/8\pi$ being closer to either 0 or 1, which is also apparent from the different curves plotted in Fig.~\ref{fig:ebetaphi_XYZ_QEC} (top right panel). 
The corresponding correlation lengths are shown in the bottom left and bottom right panels. We do not go beyond $E_{J_0}/2E_{C_0}=0.1$. This is because for this choice, the correlation is already only a few lattice sites for both models. Thus, beyond this point, the field theory predictions are not expected to describe either lattice model very well. Note that beyond $E_{J_0}/E_{C_0}> 10^{-2}$, the correlation length of the XYZ chain actually goes up, while that in the QEC model keeps going down. The increase in $\xi_{\rm XYZ}$ for this range of $E_{J_0}/E_{C_0}$ can be understood by noticing that in this case, $|J_x - J_y|\geq|J_x - J_z|$ and the model is no longer in the regime $J_x>J_y\geq|J_z|$ [see below Eq.~\eqref{ham_xyz}]. Recalling that the phase-diagram of the XYZ chain is symmetric under permutation of the coupling constants~\cite{Baxter2013}, we can interchange $J_y$ and $J_z$ and arrive at a corresponding formula for $\xi_{\rm XYZ}$ [Eq.~\eqref{xi_xyz}] that is consistent with this behavior. Thus, in this regime, the correspondence between the qSG and the XYZ model presented in Sec.~\ref{xyz_model} is no longer valid. To check that we are not encountering any numerical artifacts of the DMRG simulations, we provide an extra check by performing computations of the entanglement spectrum of the XYZ chain and comparing with analytical predictions also in the regime $E_{J_0}/E_{C_0}> 10^{-2}$ (see below). Note that there is no such restriction on the parameter space for the QEC model since the two qSG parameters, $\beta, M_0$, can be {\it independently} controlled by tuning corresponding lattice parameters $E_J, E_{J_0}$. 

\section{Entanglement spectrum of the qSG model}
\label{ent_prop}
In the previous section, we have computed various thermodynamic quantities of the qSG model, namely, the one and two-point functions of vertex operators with the QEC regularization and discussed how the scaling regime is reached in comparison to the XYZ chain. Now, we compute the entanglement spectrum of the qSG model. 

\begin{figure}
  \centering
  \includegraphics[width = 0.95\textwidth]{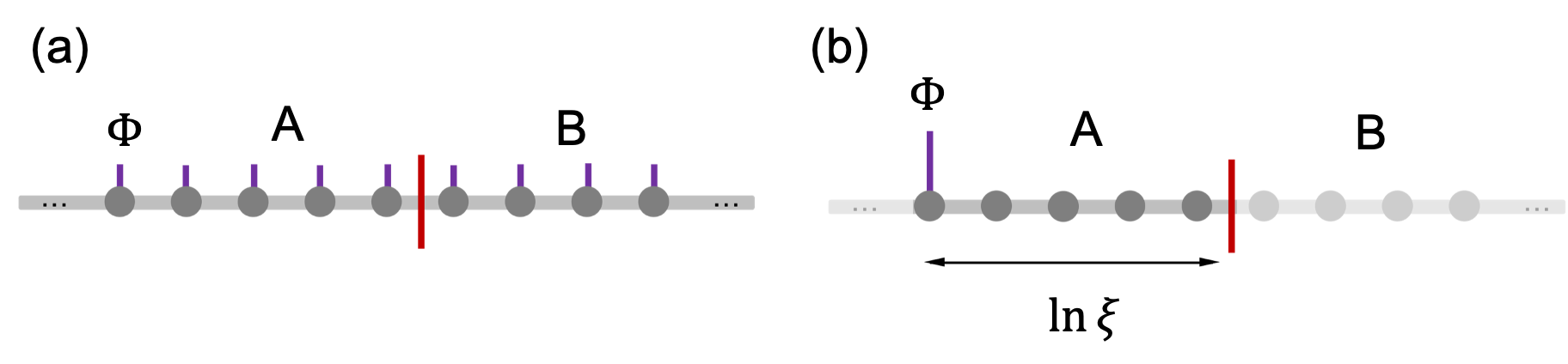}
  \caption{\label{fig:bulk_boundary} Correspondence between the entanglement spectrum of a CFT perturbed by a single primary field $\Phi$ [panel (a)] and the physical spectrum of a boundary CFT [panel (b)]. The perturbation $\Phi$ is denoted by the purple lines. The boundary CFT has the following boundary conditions: free boundary condition arising from the entanglement cut on one end and a boundary field $\Phi$ at the other. Note that the boundary CFT is defined over a length $\ln\xi$, where $\xi$ is the correlation length.  }
\end{figure}

First, we briefly summarize the generic behavior of the entanglement spectrum for a partitioning of an infinite system into two halves for massive field theories following Ref.~\cite{Cho2017}. Consider a CFT perturbed by a single primary field $\Phi$ (the generalization to multiple perturbations is straightforward). The entanglement spectrum for this massive theory is given by the physical spectrum of a corresponding boundary CFT over a length interval $\ln\xi$, where $\xi$ is the (finite) correlation length~\cite{Cho2017}. The two spectra are equal up to rescalings and overall shifts and comprise equidistant levels. The boundary CFT has two boundary conditions: free boundary condition at one end, which arises from the entanglement cut and a boundary field, $\Phi$, at the other end (see Fig.~\ref{fig:bulk_boundary}). The equality of the two spectra is correct up to exponential corrections. This leads to a restriction on the number of low-lying entanglement energy levels which are in correspondence with the boundary CFT spectrum~\cite{Cho2017}. Note that when the system is critical {\it i.e.,} $\Phi=0$ and $\xi\rightarrow\infty$, the second boundary condition (that on the left end in  Fig.~\ref{fig:bulk_boundary}) is inherited from the original model whose entanglement spectrum is being computed~\cite{Roy2020a,Cardy2016}. 

The aforementioned relationship between the two spectra holds independently of whether the perturbation, $\Phi$, is integrable or not. But, the qSG model is an integrable deformation of the free, compactified boson CFT and thus, the question arises as to whether one can use integrability to glean additional information about the qSG entanglement spectrum. To that end, we can use the fact that the qSG model arises as the scaling limit of the quantum XYZ chain (Sec.~\ref{xyz_model}). The latter spin-chain or its ``classical" version, the eight-vertex model~\cite{Baxter2013}, falls within the category of integrable lattice models which exhibit equidistant levels for the {\it entire} entanglement spectra -- notable other examples include the transverse-field Ising and the XXZ models~\cite{Peschel1991, Peschel1999, Calabrese2010}. We compute the entanglement spectrum of the XYZ chain both analytically and numerically in Sec.~\ref{xyz_es}. In particular, we show that the level-spacing, denoted by $\varepsilon_{\rm XYZ}$, goes as $1/\ln\xi_{\rm XYZ}$, where $\xi_{\rm XYZ}$ is the correlation length of the XYZ model [see Eq.~\eqref{xi_xyz}] as long as the system size is much larger than $\xi_{\rm XYZ}$. This behavior is consistent with what is predicted in Ref.~\cite{Cho2017}. 

After computing the qSG entanglement spectrum using the XYZ chain, we compute the same for the QEC lattice model using DMRG. The primary motivation for this computation is to investigate to what extent the low-lying entanglement spectrum of the XYZ chain is a universal feature of the qSG model. As will be shown in Sec.~\ref{qec_es}, the spectra computed using the XYZ and the QEC lattices have identical degeneracies. Furthermore, the level spacing of the entanglement spectrum computed using the QEC lattice model, denoted by $\varepsilon_{\rm QEC}$ also goes as $1/\ln\xi_{\rm QEC}$, where $\xi_{\rm QEC}$ is the correlation length fo the QEC lattice model. Without fine-tuning of the microscopic models, there is no reason why $\varepsilon_{\rm XYZ}$ would be equal to $\varepsilon_{\rm QEC}$. However, both quantities scale inversely with the logarithm of the respective correlation lengths. At the same time, both $\xi_{\rm XYZ}$ and $\xi_{\rm QEC}$ depend on the qSG mass-parameter, $M_0$, as $M_0^{-1/(2-\beta^2/4\pi)}$. Thus, from purely dimensional considerations, we can conclude that $\varepsilon_{\rm XYZ}$ and $\varepsilon_{\rm QEC}$ are linearly dependent on each other: 
\begin{equation}
\label{qec_xyz_dep}
\varepsilon_{\rm QEC}(\beta, M_0) = a_0(\beta) + a_1(\beta) \varepsilon_{\rm XYZ}(\beta, M_0),
\end{equation}
where $a_0(\beta), a_1(\beta)$ are (possibly non-universal) functions that depend on the parameters of the two lattice models. Here, we have also explicitly indicated the dependence of the entanglement level spacings on the qSG parameters: $\beta, M_0$. We verify this linear dependence in Sec.~\ref{qec_es}. The secondary motivation for this computation is to demonstrate that the QEC lattice model continues to provide meaningful prediction for the qSG entanglement spectrum even when the XYZ chain is no longer in the scaling limit (see discussion of Sec.~\ref{scale_correction}). 

\subsection{XYZ model}
\label{xyz_es}
Now, we compute the qSG entanglement spectrum using the XYZ chain. The XYZ Hamiltonian [see Eq.~\eqref{ham_xyz}] can be related to the transfer-matrix of the classical eight-vertex model. To establish this relation, it is useful to consider the principal regime for the two models~\cite{Baxter2013}. Denote the XYZ couplings in the principal regime by $J_\alpha^p$, where $\alpha = x,y,z$. Then, the principal regime is given by 
\begin{equation}
  \label{princ_XYZ}
  |J_y^p|\leq J_x^p\leq - J_z^p.
\end{equation}
The couplings of the XYZ chain can be related to the two parameters of the classical eight-vertex model, denoted by $\Gamma, \Delta$. In the principal regime, they are given by 
\begin{equation}
  \Gamma_p = \frac{J_y^p}{J_x^p}, \Delta_p = \frac{J_z^p}{J_x^p}.
  \label{XYZ_SG_coupling}
\end{equation}
As a result, in the principal regime, for the eight-vertex model, 
\begin{equation}
  |\Gamma_p|\leq 1, \Delta_p \leq -1. 
  \label{princ_8_vert}
\end{equation}
Since we are interested in the entanglement properties of the XYZ chain, we need two further parameters, $\lambda, k$, given by~\cite{Baxter2013}
\begin{align}
  \label{k_def}
  \frac{2\sqrt{k}}{1+k} = \sqrt{\frac{1-\Gamma_p^2}{\Delta_p^2 - \Gamma_p^2}},\  -i{\rm sn}(i\lambda, k) = \frac{1}{\sqrt{k}}\sqrt{\frac{1-\Gamma_p}{1+\Gamma_p}},
\end{align}
where ${\rm sn}$ is the Jacobi sine function~\cite{Baxter2013}.  Here, $0\leq k\leq1$ and $0\leq\lambda\leq I(k')$, where $k' = \sqrt{1-k^2}$ and $I(k)$ is the complete elliptic integral of the first kind with modulus $k$.  
The entanglement spectrum for the XYZ chain can be related to the spectrum of the CTM of the eight-vertex model~\cite{Ercolessi2009, Evangelisti2013}. The entanglement spectrum comprises equidistant levels, with the level spacing given by $\varepsilon_{\rm XYZ}$, given by~\cite{Baxter2013}
\begin{equation}
  \varepsilon_{\rm XYZ} = \frac{\pi \lambda}{I(k)}. 
  \label{epsilon_def}
\end{equation}
\begin{figure}
  \centering
  \includegraphics[width = 0.75\textwidth]{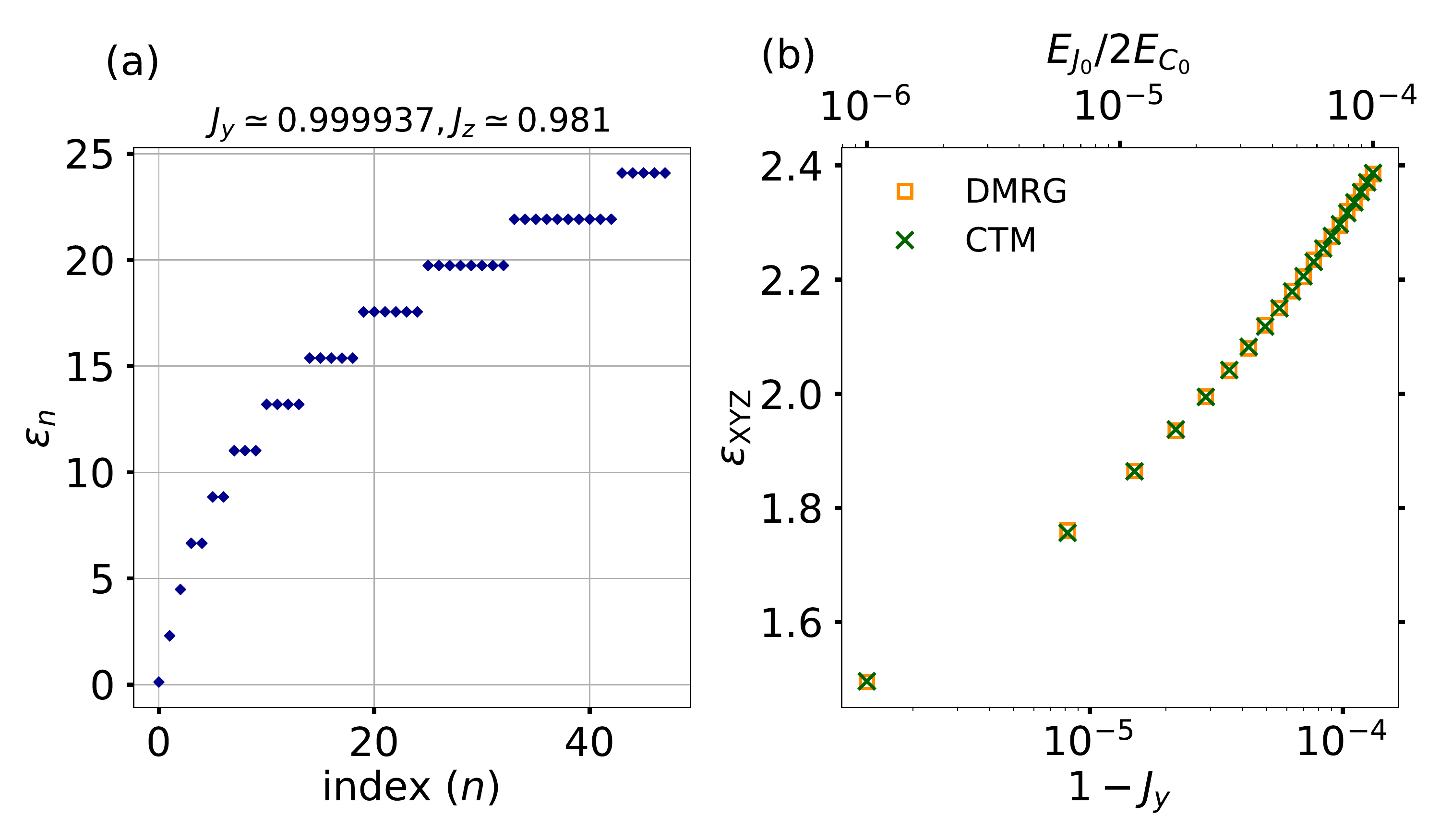}
  \caption{\label{fig:XYZ_es} (a) Entanglement spectrum for the XYZ chain obtained with infinite DMRG. We chose $\beta^2/8\pi \simeq 0.063$, which corresponds to $J_z\simeq0.981$~[see Eq.~\eqref{xi_xyz}] and $J_y\simeq0.999937$. The spectrum $\{\varepsilon_n\}$, comprises equidistant levels. The level spacing ($\varepsilon_{\rm XYZ}$) is obtained by extracting the slope by performing a linear fit for the first three eigenvalues. (b) Extracted values of $\varepsilon_{\rm XYZ}$ as $J_y$ is varied. The infinite DMRG results are shown in empty squares, while the CTM results obtained by using Eqs.~(\ref{k_def}, \ref{epsilon_def}) are shown with crosses. For reference, we show the corresponding value of the qSG mass-parameter in terms of the QEC circuit-parameter [see Eqs.~\ref{S_SG_1}, \ref{beta_def}] on the top x-axis. }
\end{figure}

\begin{figure}
  \centering
  \includegraphics[width = 0.9\textwidth]{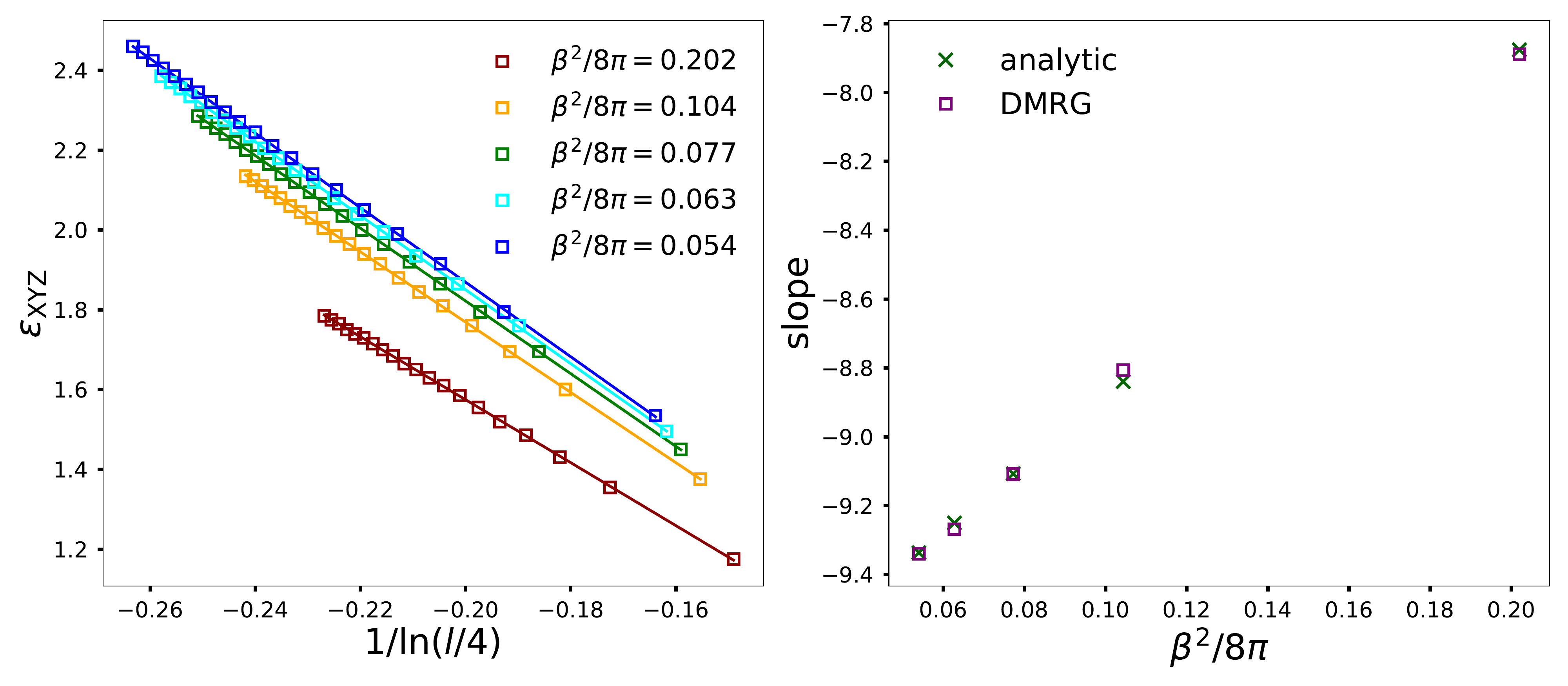}
  \caption{\label{fig:eps_l} (Left) Variation of the entanglement spectrum spacing, $\varepsilon_{\rm XYZ}$, as function of $l$ for different $\beta^2/8\pi$. As expected from Eq.~\eqref{eps_l}, $\varepsilon_{\rm XYZ}$ exhibits a linear dependence with $1/\ln(l/4)$. (Right) The slopes extracted from panel (a) as a function of $\beta^2/8\pi$, together with the analytical values, given by $-\pi^2(1-\beta^2/8\pi)$, are shown.}
\end{figure}

Now, we compute the scaling of the level-spacing of the entanglement spectrum, $\varepsilon_{\rm XYZ}$, as the mass-parameter of the sine-Gordon action, $M_0$ of Eq.~\eqref{S_SG}, is taken to zero. 
This corresponds to taking the limit $J_y/J_x\rightarrow1^{-}$ in the XYZ chain. For the purposes of the calculation, we set $J_x=1$ and consider the case when $|J_y/J_z|\geq1$ (the other case can be analyzed similarly). Our goal is to compute the scaling of the level-spacing of the entanglement spectrum, $\varepsilon_{\rm XYZ}$, given in Eq.~\eqref{epsilon_def} as $J_y/J_x\rightarrow1^{-}$. To that end, we first define the couplings in the principal regime: 
\begin{align}
  J_x^p = J_y, \ J_y^p = - J_z, \ J_z^p = -J_x = -1. 
  \label{Jps}
\end{align}
Thus, the limit $J_y/J_x\rightarrow 1^-$ is equivalent to $J_x^p/J_z^p\rightarrow -1^+$, which in turn implies $\Delta_p\rightarrow -1^-$ [see Eq.~\eqref{XYZ_SG_coupling}]. From Eq.~\eqref{k_def}, this implies $k\rightarrow 1^-$. We will also use the fact that
\begin{equation}
  -i{\rm sn}(i\lambda,k)\rightarrow\tan\lambda, \ I(k)\rightarrow -\frac{1}{2}\ln\frac{1-k}{8}
\end{equation}
as $k\rightarrow1^-$. To quantify deviations from the critical point, we define two small parameters:
\begin{align}
  \delta \equiv 1-k,\ x \equiv 1 - J_y.
  \label{delta_x_def}
\end{align}
Then, from Eq.~\eqref{k_def}, we get 
\begin{align}
\label{delta_eqn}
  \delta &= \frac{2\sqrt{2x}}{\sin(\beta^2/8)} - \frac{4x}{\sin^2(\beta^2/8)}\\
  \lambda &= \tan^{-1}\Bigg[\frac{1}{1-\delta}\sqrt{\frac{1-\Gamma_p}{1+\Gamma_p}}\Bigg] = \frac{\pi}{2}\Bigg(1-\frac{\beta^2}{8\pi}\Bigg) + \frac{\sqrt{x}}{2},
	  \label{lambda_eq}
\end{align}
where we have kept terms up to ${\cal O}(x)$. Next, we use 
\begin{equation}
  l^2 \simeq \frac{2x}{\sin^2(\beta^2/8)}
  \label{l_k_1}
\end{equation}
as $k\rightarrow 1^-$ to get
\begin{align}
  \varepsilon_{\rm XYZ} &\simeq -\frac{\pi^2}{\ln(l/4)}\Bigg(1-\frac{\beta^2}{8\pi}\Bigg) + {\cal O}\Bigg(\frac{1}{\ln^2 l}\Bigg),
  \label{eps_l}
\end{align}
which is commensurate with the general statement that the entanglement spectrum gap closes as $1/\ln\xi_{\rm XYZ}$ to leading order~\cite{Cho2017, Calabrese2010}, see Eq.~\eqref	{xi_xyz}.
We check the leading order term by analyzing the dependence of $\varepsilon_{\rm XYZ}$ as a function of $l$. This is shown in Fig.~\ref{fig:eps_l}, which confirms the predicted linear dependence with $1/\ln(l/4)$ with a slope that is close to $-\pi^2(1-\beta^2/8\pi)$. 

\subsection{QEC model}
\label{qec_es}
\begin{figure}
  \centering
  \includegraphics[width = \textwidth]{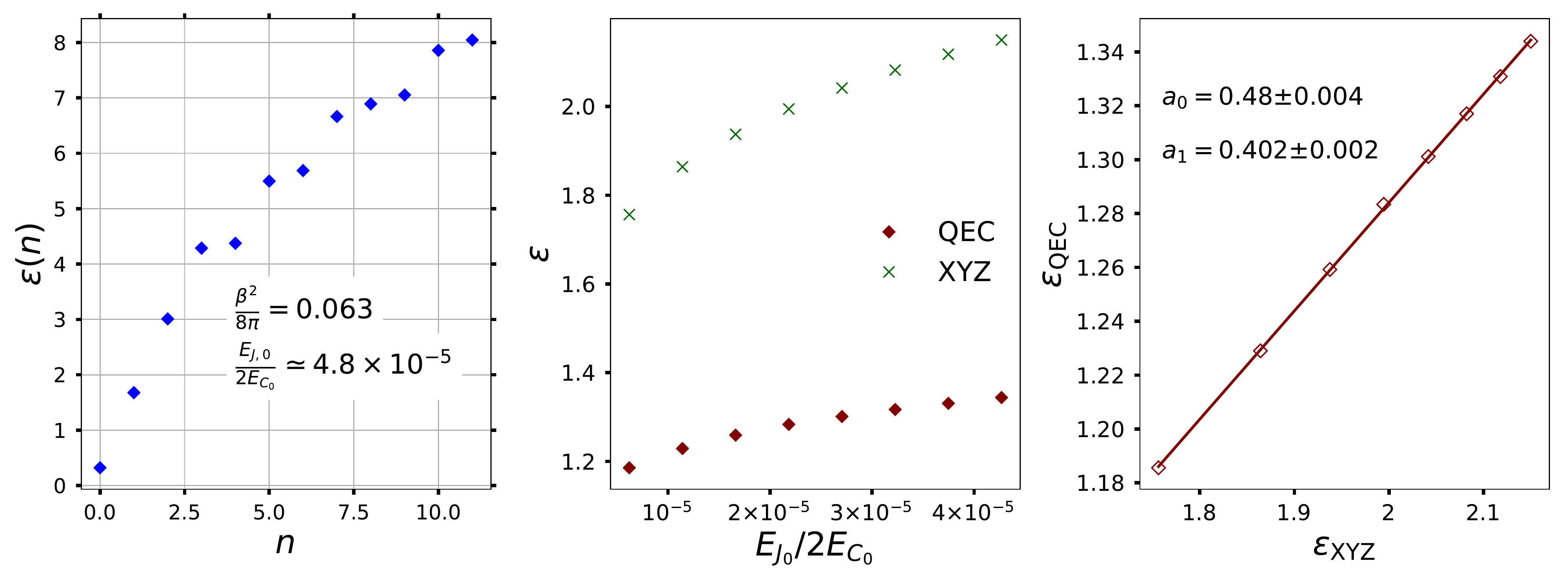}
  \caption{\label{fig:es_qec_xyz} Entanglement spectrum properties for the QEC lattice model using infinite DMRG. We chose $\beta^2/8\pi\simeq0.063$. (Left) First 12 entanglement energies computed for the QEC model. We chose $E_{J_0}/E_{C_0}\simeq4.8\times10^{-5}$. The degeneracies for the plotted levels are the same as for the XYZ chain [see panel (a) of Fig.~\ref{fig:XYZ_es} for the corresponding results]. Note that the finite size effects are larger for the QEC lattice compared to the XYZ chain (see main text for discussion). (Center) The entanglement spectrum level spacings, $\varepsilon_{\rm XYZ}$ and $\varepsilon_{\rm QEC}$, as a function of $E_{J_0}/2E_{C_0}$. As argued earlier, the two level-spacings are not equal for each parameter choice, but are linearly dependent on each other.  (Right) Variation of $\varepsilon_{\rm QEC}$ as a function of the $\varepsilon_{\rm XYZ}$, verifying the linear dependence argued in Sec.~\ref{ent_prop}. The coefficients of the linear fit and the errors in their determination are shown [see Eq.~\eqref{qec_xyz_dep}]. }
\end{figure}

In this section, we compute the qSG entanglement spectrum using the QEC lattice model. As discussed earlier, the low-lying part of the spectrum should be a characteristic of the qSG field theory and thus, should be comparable to the results obtained from the XYZ chain (see in Fig.~\ref{fig:XYZ_es}). The results are shown in Fig.~\ref{fig:es_qec_xyz} for $\beta^2/8\pi\simeq0.063$ (similar results were obtained for other choices and are not shown for brevity). As seen from the left panel, the low-lying entanglement spectrum exhibits the expected equidistant level structure, with the same degeneracy structure given in Fig.~\ref{fig:XYZ_es}~(a). Despite the overall degeneracy structure being consistent, it is clear that data quality for the QEC model is worse compared to the XYZ chain. One of the reasons for this is that the XYZ chain, unlike the QEC model, exhibits the equidistant structure for the {\it entire} entanglement spectrum due to its relationship to the eight-vertex model. We are not aware of any such deep connections for the QEC model. We believe the worse data quality is also partially due to the non-universal lattice effects which affect the two models differently. Finally, at a more pragmatic level, the large local Hilbert space of the QEC model makes the computations much more resource-consuming compared to the XYZ chain. This restricts the size of the bond-dimensions that are accessible for a moderate-scale simulation effort pursued in this work. 
In the center panel, we show the level-spacings for different choices of $E_{J_0}/E_{C_0}$ as obtained from the QEC and the XYZ models for $\beta^2/8\pi =0.063$. As argued earlier [see discussion before Eq.~\eqref{qec_xyz_dep}], the two level-spacings are not equal to each other, but are linearly related. The verification of this linear dependence is shown in the right panel of Fig.~\ref{fig:es_qec_xyz}.

\begin{figure}
  \centering
  \includegraphics[width = 0.9\textwidth]{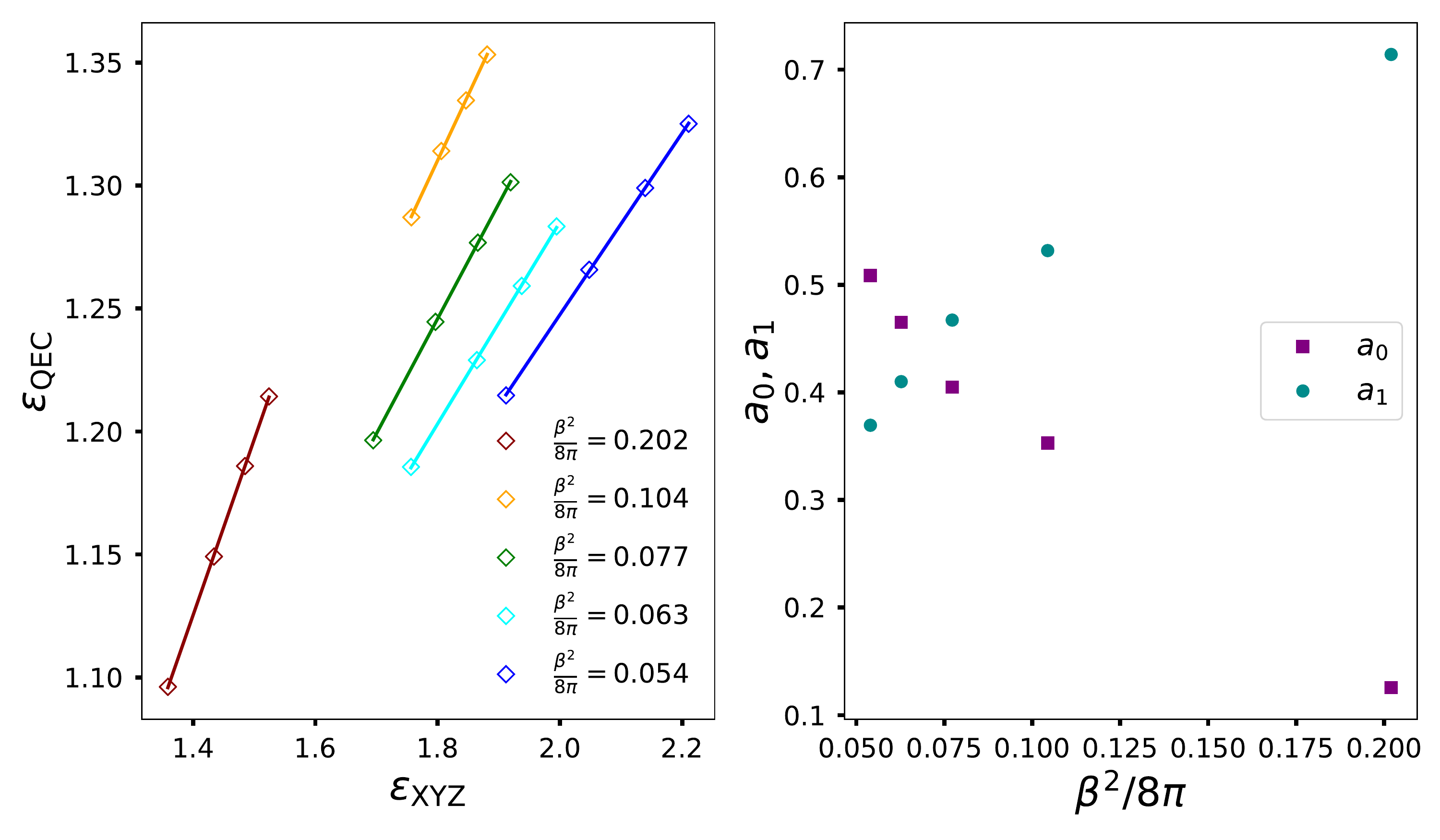}
  \caption{\label{fig:es_qec_xyz_2} (Left) Verification of the linear dependence of $\varepsilon_{\rm QEC}$ on $\varepsilon_{\rm XYZ}$ for different choices of $\beta^2/8\pi$.  (Right) The parameters of the linear fit, $a_0, a_1$ [see Eq.~\eqref{qec_xyz_dep}] as a function of $\beta^2/8\pi$. }
\end{figure}

Next, in Fig.~\ref{fig:es_qec_xyz_2}~(left panel), we verify the linear dependence of $\varepsilon_{\rm QEC}$ on $\varepsilon_{\rm XYZ}$ for different choices of $\beta^2/8\pi$. In the right panel, the corresponding parameters of the linear dependencies, given in Eq.~\eqref{qec_xyz_dep}, are shown. 
At this point, we do not have a deep understanding of the $a_0, a_1$ and their functional dependency on $\beta$. It is plausible that these are non-universal functions of $\beta$ which depend on the details of the QEC and XYZ lattice models, but we leave a detailed investigation for a future work. 

\subsection{Corrections to scaling in the entanglement spectrum: QEC vs XYZ}
\label{es_qec_vs_xyz}
\begin{figure}
  \centering
  \includegraphics[width = 0.55\textwidth]{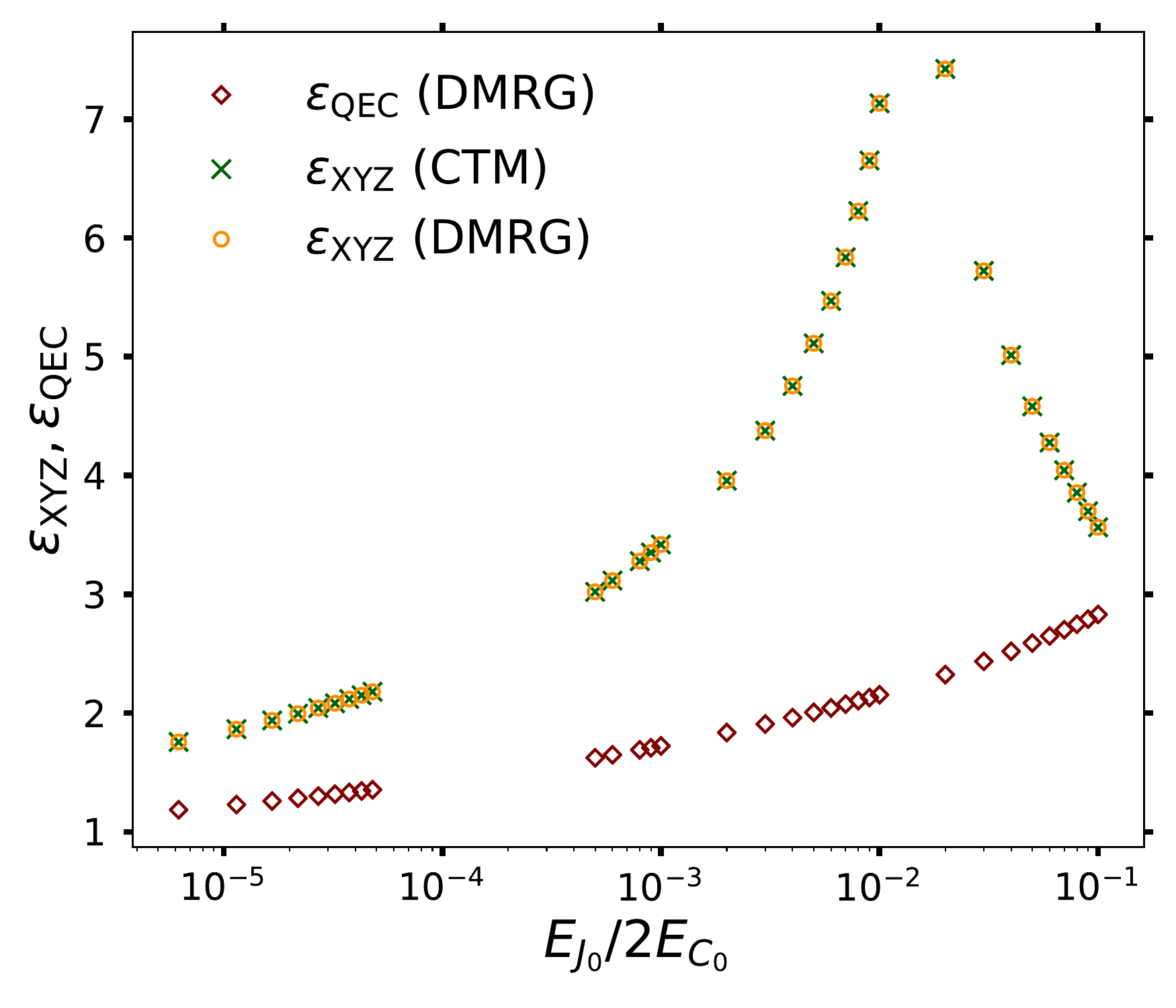}
  \caption{\label{fig:es_qec_xyz_3} Level-spacing of the qSG entanglement spectrum as computed with the QEC and the XYZ lattice models. We chose $\beta^2/8\pi\simeq0.063$ (similar results were obtained for other choices and are not shown for brevity). For both models, we show the DMRG results. Furthermore, for the XYZ chain, we show the analytic CTM predictions (see Sec.~\ref{xyz_es}) as well. For $E_{J_0}/E_{C_0}< 10^{-3}$, both the models give physically meaningful predictions, with $\varepsilon_{\rm QEC}$ being linearly dependent on $\varepsilon_{\rm XYZ}$. However, further increase of $E_{J_0}/E_{C_0}$ causes $\varepsilon_{\rm XYZ}$ to increase much faster compared to $\varepsilon_{\rm QEC}$. Beyond $E_{J_0}/E_{C_0}>10^{-2}$, $\varepsilon_{\rm XYZ}$ goes down, which is clearly incompatible with the expectation that the entanglement level-spacing increases with increase of the qSG mass-parameter. The results for the QEC model, however, continue to provide physically meaningful predictions, increasing steadily with $E_{J_0}/E_{C_0}$. The perfect overlap of the CTM and the DMRG results for the XYZ chain show that what we are observing is not a numerical artifact of DMRG; rather, it is the qSG-XYZ correspondence that is no longer valid for $E_{J_0}/E_{C_0}>10^{-3}$ (see Secs.~\ref{scale_correction}, \ref{es_qec_vs_xyz} for more discussion). }
\end{figure}

Here, we demonstrate that the corrections to scaling, which are noticeable for $E_{J_0}/E_{C_0}>10^{-3}$ and lead to incorrect dependence of the qSG vertex operator $e^{i\beta\phi/2}$ in the XYZ chain (see Sec.~\ref{scale_correction}), also causes the qSG entanglement spectrum to be incorrectly inferred from the XYZ results. This is in contrast to the QEC model, which continues to provide meaningful physical predictions for the entanglement spectrum for these choices of $E_{J_0}/E_{C_0}$. The results are shown in Fig.~\ref{fig:es_qec_xyz_3} for $\beta^2/8\pi\simeq0.063$ (similar results were obtained for other values and are not shown for brevity). For $E_{J_0}/2E_{C_0}< 10^{-3}$, both XYZ and the QEC models provide meaningful results for the qSG entanglement spectrum and the level-spacings, $\varepsilon_{\rm QEC}$ and $\varepsilon_{\rm XYZ}$, are linearly dependent on each other (see Fig.~\ref{fig:es_qec_xyz_2}). However, upon further increase of $E_{J_0}/E_{C_0}$, $\varepsilon_{\rm XYZ}$ grows much faster violating the linear dependence. The point of departure of the linear dependence coincides precisely with the departure of the linear dependence of the vertex operator $e^{i\beta\phi/2}\sim\sigma^+$ in Fig.~\ref{fig:ebetaphi_XYZ_QEC}. Beyond $E_{J_0}/2E_{C_0}\sim10^{-2}$, $\varepsilon_{\rm XYZ}$ actually decreases. Clearly, for these parameters, $\varepsilon_{\rm XYZ}$ cannot correspond to the level-spacing of the qSG entanglement spectrum. This is because increasing $E_{J_0}/E_{C_0}$ increases the mass-gap of the qSG model, which would increase the entanglement-spectrum level spacing of the qSG model. Comparing to Fig.~\ref{fig:ebetaphi_XYZ_QEC} (bottom right panel), this region of decreasing $\varepsilon_{\rm XYZ}$ corresponds precisely to the region where the correlation length of the XYZ chain, $\xi_{\rm XYZ}$, increases. We again emphasize that the physics of the XYZ model is perfectly consistent -- increasing correlation length should be associated with a decreasing entanglement level-spacing. An independent check of this is provided by the CTM calculations of the entanglement spectrum (see Sec.~\ref{xyz_es}). As seen from Fig.~\ref{fig:es_qec_xyz_3}, the agreement between the DMRG and the CTM calculations is excellent for {\it all} choices of $E_{J_0}/E_{C_0}$. Finally, the results obtained using the QEC lattice exhibit the expected behavior for all choices of $E_{J_0}/E_{C_0}$. 

Before concluding this section, we point out that the qSG entanglement spectrum could, in principle, be inferred using the spectrum of the boundary sine-Gordon model~\cite{Bajnok2002, Ghoshal1994, Fendley1994} after taking the strength of the bulk perturbation of the latter model to zero. Potentially, this could also be a way to investigate the fate of the boundary bound-states of the boundary sine-Gordon model in the massless bulk limit and we hope to return to this problem in the future. We did check that the degeneracies of the entanglement spectrum of the qSG do indeed match that of the free, compactified boson CFT with Dirichlet boundary conditions~\cite{Roy2020a}. The latter boundary condition can be viewed as an extreme case when the strength of the boundary potential is taken to infinity. 

\section{Summary and outlook}
\label{concl}
To summarize, we numerically analyzed with DMRG a faithful, analog, quantum simulator built with QEC elements for the qSG model in 1+1 space-time dimensions. 
The QEC model provides a lattice-regularization of the qSG model using local interactions that can be physically realized by a straightforward generalization of current experimental works. By computing various zero-temperature  thermodynamic properties of the QEC model with DMRG and comparing with the qSG field theory predictions, we numerically demonstrate that the QEC lattice indeed realizes the qSG model. Furthermore, we show that in contrast to the integrable XYZ-chain regularization, the QEC lattice model is less susceptible to corrections to scaling. In contrast to the XYZ chain, where the spin-operator $\sigma^+$ corresponds to the qSG vertex operator $e^{i\beta\phi/2}$, the QEC model directly starts with lattice versions of the operators $e^{i\beta\phi}$. Furthermore, we computed the entanglement spectrum of the qSG model using both the XYZ and the QEC models and showed that the low-lying entanglement energy levels exhibit the same set of degeneracies. We provided a scaling argument to show that the level-spacings of the low-lying entanglement spectrum for the two models are linearly related to each other and verified this claim with numerical results. Finally, we also showed that in the XYZ chain, the same corrections to scaling that plague the correlation functions of the vertex operators also cause the model to predict unphysical values of the qSG entanglement spectrum. The latter problem is also remedied by the QEC lattice model. 

The current work gives rise to many, new, potentially-fruitful research directions and we discuss some of them below. First, concerning the qSG model, an experimentally-realizable, numerically-tractable lattice model potentially opens the door to the investigation of several open problems -- examples include many finite temperature properties of the correlation functions for the qSG model~\cite{Essler2009, Szecsenyi2013, Buccheri2014}. At the same time, this work opens the possibility of experimentally exploring non-equilibrium phenomena in the qSG model, which, in the recent years, have received enormous interest~\cite{Bertini2014,Kormos2016, Cort_s_Cubero2017, Horvath2018, Horvath2019, Bertini2019, Rylands2019, Kukuljan2019, Rylands2020}. 
Second, our approach to faithfully simulate QFTs with QECs can be used to investigate multi-field models. For the latter, it is crucial that the underlying lattice degrees of freedom faithfully give rise to the continuum ones, without resorting to mathematical manipulations like bosonization. This is because properties as fundamental to the QFT as integrability can be difficult to relate in the fermionic and bosonic counterparts -- {\it e.g.}, the quantum double sine-Gordon model, which can be faithfully realized with QECs~\cite{Roy2019}. We aim to analyze the latter model with QECs in the near future. Third, the analog QEC simulator for the qSG model analyzed in this work can be readily generalized to include integrability-breaking perturbations~\cite{Roy2019}. After all, the eventual goal is to simulate interacting QFTs to answer questions which are intractable with analytical methods. This is possible with QEC lattices, which can be used to simulate the massive Schwinger model~\cite{ZinnJustin2002} or a two-frequency generalization of the qSG model~\cite{Bajnok2000, Mussardo2004}. In fact, the basic primitives for realizing the two-frequency generalization have already been experimentally demonstrated~\cite{Gladchenko2009, Bell2014}. 
Fourth, analog QEC simulation of QFTs can also be implemented in higher dimensions. In particular, QEC arrays in 2+1 space-time dimensions have already been built and analyzed experimentally in the context of realizing interacting bosonic models, even in hyperbolic space~\cite{Kollar2019}. Given this recent experimental progress, we are optimistic of the use of QECs for investigation of 2+1D QFTs in the near future.

\section{Acknowledgments}
AR acknowledges discussions with Ian Affleck, Pasquale Calabrese, Michel Devoret, Robert Konik, Ramamurti Shankar and A. Douglas Stone. AR is particularly grateful to Sergei Lukyanov for a critical reading of the manuscript and several important comments. FP and AR are funded by the European Research Council (ERC) under the European Unions Horizon 2020 research and innovation program (grant agreement No. 771537). FP acknowledges the support of the DFG Research Unit FOR 1807 through grants no. PO 1370/2-1, TRR80, and the Deutsche Forschungsgemeinschaft (DFG, German Research Foundation) under Germany’s Excellence Strategy EXC-2111-390814868. DS is part of the D-ITP consortium, a program of the Netherlands Organisation for Scientific Research (NWO) that is funded by the Dutch Ministry of Education, Culture and Science (OCW).  JH was supported by the U.S. Department of Energy, Office of Science, Office of Basic Energy Sciences, Materials Sciences and Engineering Division under Contract No. DE-AC02-05- CH11231 through the Scientific Discovery through Advanced Computing (SciDAC) program (KC23DAC Topological and Correlated Matter via Tensor Networks and Quantum Monte Carlo). HS was supported in part by the Advanced ERC NuQFT. 

\appendix

\section{Form-factors calculation of the two-point correlation function}
\label{two_pt_fn}
In this appendix we derive analytic results for the two-point correlation function $\langle e^{i\varphi_i}e^{-i\varphi_{i+r}}\rangle$ shown in Fig.~\ref{fig:correln}. More precisely we calculate the static, zero-temperature two-point function $\langle e^{i\beta\phi(0)}e^{-i\beta\phi(r)}\rangle$ via a form-factor expansion~\cite{Smirnov1992} directly in the continuum model \eqref{S_SG}. The basic idea is to insert a resolution of the identity between the operators, where the sum runs over all possible intermediate states. Since the spectrum of the qSG model is exactly known, these intermediate states can be classified by their particle contents (solitons, antisolitons and breathers of type $n$) and the respective momenta of the particles. Since the masses of the particles in the intermediate state will lead to an exponential decay at large distances, the leading behavior will be governed by the lightest particles. Here we consider the vacuum state $|0\rangle$, single breathers of type 1 and 2, $|\theta\rangle_{1,2}$, and two 1-breather states $|\theta_1,\theta_2\rangle_{1,1}$ respectively, where we parametize the momenta of the intermediate $n^\text{th}$ breather via their rapidities, $P=\frac{m_n}{u} \sinh\theta$. Thus we evaluate 
\begin{equation}
\begin{split}
\langle e^{i\beta\phi(0)}e^{-i\beta\phi(r)}\rangle=&\big|\langle 0|e^{i\beta\phi}|0\rangle\big|^2
+\int\frac{d\theta}{2\pi}\big|\langle 0|e^{i\beta\phi}|\theta\rangle_1\big|^2e^{-i\frac{m_1}{u}r\sinh\theta}
+\int\frac{d\theta}{2\pi}\big|\langle 0|e^{i\beta\phi}|\theta\rangle_2\big|^2e^{-i\frac{m_2}{u}r\sinh\theta}\\
&+\frac{1}{2}\int\frac{d\theta_1d\theta_2}{(2\pi)^2}\big|\langle 0|e^{i\beta\phi}|\theta_1,\theta_2\rangle_{1,1}\big|^2e^{-i\frac{m_1}{u}r\sum_i\sinh\theta_i}+\ldots
\end{split}
\end{equation}
where the factor $\frac{1}{2}$ in the last term avoids double counting, and the dots represent terms with heavier intermediate states. For the parameters of Fig.~\ref{fig:correln}, $\beta^2/8\pi=0.063$, the next terms would be the single 3-breather state (mass $m_3$), the 1-breather-2-breather pair (mass $m_1+m_2$) and the three 1-breather state (mass $3m_1$). The form factors (matrix elements) appearing in the expansion are known exactly~\cite{Lukyanov1997a}. With this a straightforward calculation yields the result
\begin{equation}
\begin{split}
\langle e^{i\beta\phi(0)}e^{-i\beta\phi(r)}\rangle=\mathcal{G}_\beta^2&\left[1+\frac{\lambda^2}{\pi}K_0\left(\frac{m_1 x}{u}\right)+\frac{\lambda^4}{\pi\,|R\bigl(\ii\pi(1+\xi_{\rm SG})\bigr)|^2}\frac{\sin^2(\pi\xi_{\rm SG})}{\sin(2\pi\xi_{\rm SG})}K_0\left(\frac{m_2 x}{u}\right)\right.\\
&\left.\ +\frac{\lambda^4}{2\pi}\int\frac{\dd\theta}{2\pi}\left|\frac{\sinh\theta}{\sinh\theta-\ii\sin(\pi\xi_{\rm SG})}\frac{1}{R(\theta+\ii\pi)}\right|^2K_0\left(\frac{2m_1x}{u}\cosh\frac{\theta}{2}\right)+\ldots\right],
\end{split}
\label{eq:zerotemperature}
\end{equation}
where $\mathcal{G}_\beta$ is given in Eq.~\eqref{ebetaphi}, and the other parameters are 
\begin{eqnarray}
\lambda&=&2\cos\frac{\pi\xi_{\rm SG}}{2}\sqrt{2\sin\frac{\pi\xi_{\rm SG}}{2}}\exp\left[-\int_0^{\pi\xi_{\rm SG}}\frac{\dd t}{2\pi}\frac{t}{\sin t}\right],\\
R\bigl(\ii\pi(1+\xi_{\rm SG})\bigr)&=&\exp\left[8\int_0^\infty\frac{\dd t}{t}\frac{\sinh(t)\sinh(t\xi_{\rm SG})\sinh\bigl(t(1+\xi_{\rm SG})\bigr)}{\sinh^2(2t)}\left(\sinh^2(t\xi_{\rm SG})+\frac{1}{2}\right)\right],\quad\\
R(\theta+\ii\pi)&=&\exp\left[8\int_0^\infty\frac{\dd t}{t}\frac{\sinh(t)\sinh(t\xi_{\rm SG})\sinh\bigl(t(1+\xi_{\rm SG})\bigr)}{\sinh^2(2t)}\left(\frac{1}{2}-\sin^2\left(\frac{t\theta}{\pi}\right)\right)\right].
\end{eqnarray}
The expansion \eqref{eq:zerotemperature} contains the leading terms at large distances, with the four given terms falling off as $1$, $e^{-m_1r/u}$, $e^{-m_2r/u}$ and $e^{-2m_1r/u}$, respectively. The dots correspond to higher-order terms falling of at least as $\sim e^{-m_3r/u}$ at large distances. 

\bibliography{library_1}

\end{document}